\documentclass[%
    paper=A4,               
    twoside=true,           
    openright,              
    parskip=half,           
    chapterprefix=true,     
    11pt,                   
    headings=normal,        
    bibliography=totoc,     
    listof=totoc,           
    titlepage=on,           
    captions=tableabove,    
    chapterprefix=false,    
    appendixprefix=false,    
    draft=false,            
]{scrreprt}%

\PassOptionsToPackage{utf8}{inputenc}
\usepackage{inputenc}

\usepackage{listings}
\usepackage{tabularx}

\newcommand{\FG}[1]{Figure~\ref{#1}}

\newcommand{\thesisTitle}{How to Improve Your Virtual Experience -- Exploring the Obstacles of Mainstream VR}
\newcommand{\thesisName}{Andrey Krekhov}
\newcommand{\thesisSubject}{Cumulative Dissertation}
\newcommand{\thesisDate}{December 18, 2019}

\newcommand{\thesisFirstReviewer}{Prof. Dr. Jens Krüger}

\newcommand{\thesisSecondReviewer}{Prof. Dr. Maic Masuch}

\newcommand{\thesisUniversity}{\protect{University of Duisburg-Essen}}

\newcommand{\thesisUniversityGroup}{High Performance Computing Group (HPC)}
\newcommand{\thesisUniversityCity}{47057 Duisburg, Germany}
\newcommand{\thesisUniversityStreetAddress}{Lotharstr. 65}

\usepackage[english]{babel} 
\PassOptionsToPackage{
    figuresep=colon,%
    hangfigurecaption=false,%
    hangsection=true,%
    hangsubsection=true,%
    sansserif=false,%
    configurelistings=true,%
    colorize=full,%
    colortheme=bluemagenta,%
    configurebiblatex=true,%
    bibsys=biber,%
    bibfile=bib-refs,%
    bibstyle=alphabetic,%
    bibsorting=nty,%
}{cleanthesis}
\usepackage{cleanthesis}
\usepackage{enumitem}

\hypersetup{
    pdftitle={\thesisTitle},    
    pdfsubject={\thesisSubject},
    pdfauthor={\thesisName},    
    plainpages=false,           
    colorlinks=false,           
    pdfborder={0 0 0},          
    breaklinks=true,            
    bookmarksnumbered=true,     %
    bookmarksopen=true          %
}

\begin{document}


\renewcaptionname{english}{\figurename}{Fig.}
\renewcaptionname{english}{\tablename}{Tab.}

\pagenumbering{roman}			
\pagestyle{empty}				
%


\begin{titlepage}
	\pdfbookmark[0]{Titlepage}{Titlepage}
	\tgherosfont
	\centering


	\vspace*{50pt}
	{\LARGE \color{ctcolortitle}\textbf{\thesisTitle} \\[10mm]}
	{\Large \bigskip\bigskip\bigskip\bigskip\bigskip Von der Fakultät für Ingenieurwissenschaften, \\
	Abteilung Informatik und Angewandte Kognitionswissenschaft\\
	der Universität Duisburg-Essen\\
	\bigskip\bigskip\bigskip\bigskip
	zur Erlangung des akademischen Grades\\
	\bigskip\bigskip
	Doktor der Naturwissenschaften (Dr. rer. nat.)\\
	\bigskip\bigskip
	genehmigte kumulative Dissertation\\
	\bigskip\bigskip\bigskip
	von\\
	\bigskip\bigskip\bigskip
	Andrey Krekhov\\
	aus\\
	Ufa, Russland\\
	\bigskip\bigskip\bigskip\bigskip\bigskip\bigskip\bigskip\bigskip
	Gutachter: Prof. Dr. Jens Kr\"uger \\
	Gutachter: Prof. Dr. Maic Masuch \\
	\bigskip\bigskip\bigskip\
	Tag der mündlichen Prüfung: 18.12.2019} \\



\end{titlepage}

\hfill
\vfill
{
	\small
	\textbf{\thesisName} \\
	\textit{\thesisTitle} \\
	\thesisSubject, \thesisDate \\
	Reviewers: \thesisFirstReviewer\ and \thesisSecondReviewer \\[1.5em]
	\textbf{\thesisUniversity} \\
	\textit{\thesisUniversityGroup} \\
	\thesisUniversityStreetAddress, \thesisUniversityCity
}

\cleardoublepage

\thispagestyle{empty}
%
\pdfbookmark[0]{List of Publications}{List of Publications}
\addchap*{List of Publications}
\label{sec:publications}

\textbf{2019}

[1]: S. Cmentowski, \textbf{A. Krekhov}, and J. Kr\"uger. „"I Packed my Bag and in It I Put...": A Taxonomy of Inventory Systems for Virtual Reality Games“. In: \textit{Submission to the 2020 CHI Conference on Human Factors in Computing Systems (under review).}

[2]: \textbf{A. Krekhov} and K. Emmerich. „Player Locomotion in Virtual Reality Games“. In: \textit{The Digital Gaming Handbook (accepted).}

[3]: \textbf{A. Krekhov}, S. Cmentowski, A. Waschk, and J. Kr\"uger. „Deadeye Visualization Revisited: Investigation of Preattentiveness and Applicability in Virtual Environments“. In: \textit{IEEE Transactions on Visualization and Computer Graphics.}

[4]: \textbf{A. Krekhov}, S. Cmentowski, K. Emmerich, and J. Kr\"uger. „Beyond Human: Animals As an Escape from Stereotype Avatars in Virtual Reality Games“. In: \textit{Proceedings of the Annual Symposium on Computer-Human Interaction in Play.}

[5]: S. Cmentowski, \textbf{A. Krekhov}, and J. Kr\"uger. „Outstanding: A Multi-Perspective Travel Approach for Virtual Reality Games“. In: \textit{Proceedings of the Annual Symposium on Computer-Human Interaction in Play.}

[6]: S. Cmentowski, \textbf{A. Krekhov}, A. M\"uller, and J. Kr\"uger. „Toward a Taxonomy of Inventory Systems for Virtual Reality Games“. In: \textit{Extended Abstracts of the Annual Symposium on Computer-Human Interaction in Play Companion Extended Abstracts.}

[7]: \textbf{A. Krekhov}, S. Cmentowski, and J. Kr\"uger. „The Illusion of Animal Body Ownership and Its Potential for Virtual Reality Games“. In: \textit{2019 IEEE Conference on Games.}

[8]: \textbf{A. Krekhov}, M. Michalski, and J. Kr\"uger. „Integrating Visualization Literacy into Computer Graphics Education Using the Example of Dear Data“. In: \textit{Eurographics 2019 - Education Papers.}

[9]: S. Cmentowski, \textbf{A. Krekhov}, and J. Kr\"uger. „Outstanding: A Perspective-Switching Technique for Covering Large Distances in VR Games“. In: \textit{Extended Abstracts of the 2019 CHI Conference on Human Factors in Computing Systems.}

\textbf{2018}

[10]: \textbf{A. Krekhov} and J. Kr\"uger. „Deadeye: A Novel Preattentive Visualization Technique Based on Dichoptic Presentation“. In: \textit{IEEE Transactions on Visualization and Computer Graphics.} \textbf{Best Paper Award.}

[11]: \textbf{A. Krekhov}, S. Cmentowski, K. Emmerich, M. Masuch, and J. Kr\"uger. „GulliVR: A Walking-Oriented Technique for Navigation in Virtual Reality Games Based on Virtual Body Resizing“. In: \textit{Proceedings of the 2018 Annual Symposium on Computer-Human Interaction in Play.} \textbf{Honorable Mention Award.}

[12]: \textbf{A. Krekhov}, S. Cmentowski, and J. Kr\"uger. „VR Animals: Surreal Body Ownership in Virtual Reality Games“. In: \textit{Proceedings of the 2018 Annual Symposium on Computer-Human Interaction in Play Companion Extended Abstracts.}

\textbf{2017}

[13]: \textbf{A. Krekhov}, K. Emmerich, P. Bergmann, S. Cmentowski, and J. Kr\"uger. „Self-Transforming Controllers for Virtual Reality First Person Shooters“. In: \textit{Proceedings of the Annual Symposium on Computer-Human Interaction in Play.}

[14]: \textbf{A. Krekhov}, K. Emmerich, M. Babinski, and J. Kr\"uger. „Gestures From the Point of View of an Audience: Towards Anticipatable Interaction of Presenters With 3D Content“ In: \textit{Proceedings of the 2017 CHI
Conference on Human Factors in Computing Systems.}

\textbf{2016}

[15]: \textbf{A. Krekhov}, J. Gr\"uninger, K. Baum, D. McCann, and J. Kr\"uger. „MorphableUI: A Hypergraph-Based Approach to Distributed Multimodal Interaction for Rapid Prototyping and Changing Environments“. In: \textit{Proceedings of The 24th International Conference in Central Europe on Computer Graphics, Visualization and Computer Vision 2016.}

\textbf{2013}

[16]: F. Daiber, \textbf{A. Krekhov}, M. Speicher, J. Kr\"uger, and A. Kr\"uger. „A Framework for Prototyping and Evaluation of Sensor-based Mobile Interaction with Stereoscopic 3D“. In: \textit{ACM ITS Workshop on Interactive Surfaces for Interaction with Stereoscopic 3D.}

\cleardoublepage

\pagestyle{plain}				
%
\pdfbookmark[0]{Abstract}{Abstract}
\addchap*{Abstract}
\label{sec:abstract}

What is Virtual Reality (VR)? A professional tool, made to facilitate our everyday tasks? A conceptual mistake, accompanied by cybersickness and unsolved locomotion issues since the very beginning? Or just another source of entertainment that helps us escape from our deteriorating world? The public and scientific opinions in this respect are diverse. Furthermore, as researchers, we sometimes ask ourselves whether our work in this area is really ``worth it'', given the ambiguous prognosis regarding the future of VR. To tackle this question, we explore three different areas of VR research in this dissertation, namely locomotion, interaction, and perception. We begin our journey by structuring the VR locomotion landscape and by introducing a novel locomotion concept for large distance traveling via virtual body resizing. In the second part, we focus on our interaction possibilities in VR. We learn how to represent virtual objects via self-transforming controllers and how to store our items in VR inventories. We design comprehensive 3D gestures for the audience and provide an I/O abstraction layer to facilitate the realization and usage of such diverse interaction modalities. The third part is dedicated to the exploration of perceptual phenomena in VR. In contrast to locomotion and interaction, where we mainly deal with the shortcomings of VR, our contributions in the field of perception emphasize the strong points of immersive setups. We utilize VR to transfer the illusion of virtual body ownership to nonhumanoid avatars and exploit this phenomenon for novel gaming experiences with animals in the leading role. As one of our most significant contributions, we demonstrate how to repurpose the dichoptic presentation capability of immersive setups for preattentive zero-overhead highlighting in visualizations. We round off the dissertation by coming back to VR research in general, providing a critical assessment of our contributions and sharing our lessons learned along the way.

{
\addchap*{Abstract - German}

Was ist die virtuelle Realität (VR)? Ein professionelles Werkzeug für die alltäglichen Aufgaben? Ein konzeptioneller Fehler, begleitet von Cybersickness und von den bis heute ungelösten Problemen der Fortbewegung? Oder einfach nur eine weitere Unterhaltungsquelle, die uns die Flucht aus der unangenehmen Realität erleichtert? Die öffentliche und wissenschaftliche Meinung diesbezüglich ist gespalten. Und wir, als Forscher, fragen uns nicht selten, ob sich unsere Arbeit auf diesem Gebiet tatsächlich als lohnend herausstellt, wenn man all die zweideutigen Prognosen hinsichtlich der Zukunftsfähigkeit von VR berücksichtigt. Um eine Antwort auf diese Frage zu finden, betrachten wir in dieser Dissertation drei unterschiedliche VR-Forschungsgebiete: Lokomotion, Interaktion und Perzeption. Wir beginnen unsere Reise mit der Strukturierung der vorhandenen Fortbewegungsarten in VR und der Einführung eines neuen Konzepts zum Zurücklegen großer Distanzen durch virtuelle Körpergrößenänderung. Im zweiten Teil der Arbeit fokussieren wir uns auf die Interaktionsmöglichkeiten in VR. Wir erlernen den Bau von transformierbaren Controllern zur Repräsentation von virtuellen Objekten und betrachten das Konzept von Inventaren in VR. Ferner entwickeln wir publikumsorientierte Interaktionsgesten und führen eine Abstraktionsebene für I/O-Handling ein, um die Umsetzung und Nutzung solch mannigfaltiger Interaktionsmöglichkeiten in VR zu erleichtern. Der dritte Teil der Arbeit widmet sich der Erkundung verschiedener Wahrnehmungsphänomene in VR. Im Gegensatz zu unserer Forschung auf den Gebieten der Lokomotion und Interaktion, die sich vor allem mit den Problemen und Herausforderungen in VR beschäftigte, konzentrieren wir uns hinsichtlich Perzeption vor allem auf die starken Seiten von immersiven Setups. Wir verwenden VR, um die Illusion des virtuellen Körperbesitzes auf nicht-humanoide Avatare zu übertragen und nutzen dieses Phänomen, um neuartige Spielkonzepte mit Tieren in der Hauptrolle zu erschaffen. Als eine unserer Haupterrungenschaften zeigen wir ferner auf, wie die dichoptische Präsentationsfähigkeit eines immersiven setups ohne Zusatzaufwand für präattentives Highlighting im Bereich der Visualisierung eingesetzt werden kann. In einer abschließenden Diskussion führen wir eine kritische Bewertung unserer wissenschaftlichen Beiträge durch und teilen unsere während der Promotion erworbenen Erfahrungen auf dem Gebiet der VR-Forschung.

\cleardoublepage
\currentpdfbookmark{\contentsname}{toc}
\setcounter{tocdepth}{2}		
\tableofcontents				
\cleardoublepage

\pagenumbering{arabic}			
\setcounter{page}{1}			
\pagestyle{maincontentstyle}			


%
\chapter{Introduction}
\label{sec:intro}


Virtual Reality (VR) has taken foot in everyday life. Or did it not? On the one hand, the industry took critical leaps forward and established affordable living room VR setups. Software companies are extending their products to VR and promise unique benefits, be it for entertainment or business purposes. On the other hand, however, researchers are getting more and more vocal (cf. \FG{fig:twitter}) regarding the omnipresent obstacles of VR that have not changed that much in the last decades: limited locomotion, inferior interaction, or the risk of cybersickness, to name a few.



\begin{figure}[!h]
\centering
\includegraphics[width=1.0\columnwidth]{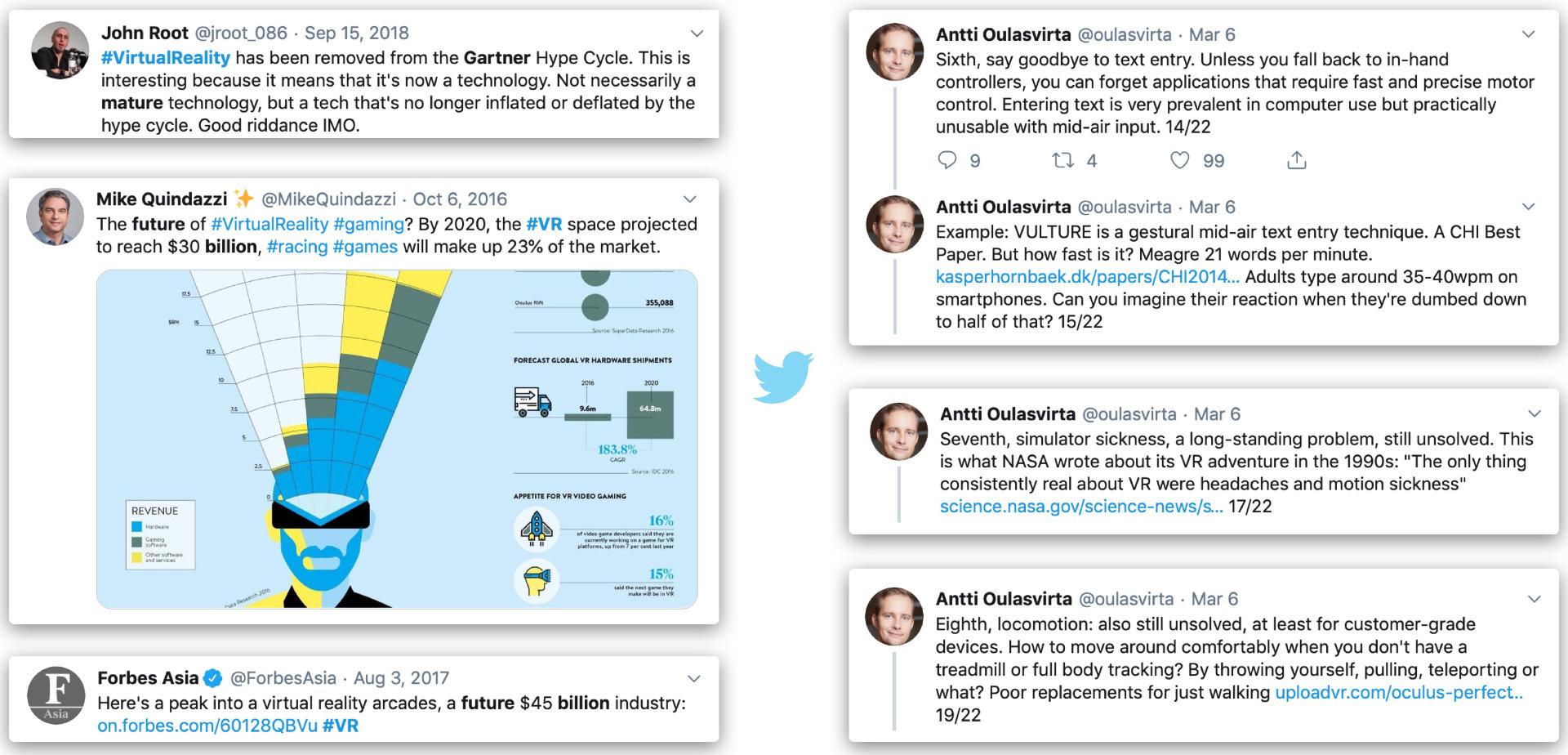}
\caption{Mixed voices on Twitter regarding the present and future of virtual reality.}
\label{fig:twitter}
\end{figure}

As prospective or even established researchers, we often question ourselves which impact our work in a certain area might have and whether it is ``worth it''. So, is VR doomed as some people claim, or are there good chances that our research efforts will be rewarded? That is exactly the central question that unites the publications gathered in this thesis. As the title already reveals, we dive into multiple VR entities to win an impression of whether and how research might change the status quo of mainstream VR. The particular contributions range from fundamental techniques, such as novel locomotion approaches, to application domains, such as scientific visualization. We dedicate the introduction section to answering three major questions of the reader: 
\begin{itemize}[noitemsep]
\item \textit{Why was the thesis written (and why should I read it)?}
\item \textit{What is the thesis (not) about?}
\item \textit{How is the manuscript organized?}
\end{itemize}
Note that this is a cumulative dissertation. Hence, in contrast to traditional manuscripts, the present synopsis is a high-level pointer to published works, rather than a verbose in-depth elaboration of a particular research topic. The main objectives of our synopsis are to establish a red thread through the conducted research, to familiarize the readers with the most important outcomes, and to draw conclusions related to the present and future of virtual reality.



\section{Motives for Revisiting VR}

One crucial argument among opponents of VR (research) is that the base approach of VR is not novel and has not evolved in the last decades. Nearly all ``VR booms'' and associated promises failed, and many of us see no reason to believe the opposite regarding the current VR era. In the early nineties, we faced first commercial setups, such as the Sega VR, and magazines predicted ``affordable VR by 1994''~\cite{engler1992affordable}. Now, more than twenty years later, VR is finally regarded as mainstream, arriving in the price range of consoles and high-end smartphones. Yet the idea remains the same: a stereo pair of images and head tracking form the basis for an experience that is well-known and being studied extensively for decades. So why is it worth reconsidering it? 

As always, the devil is in the details. While there was no major invention that revolutionized VR overnight, several subtle advances and incremental changes---both from a technical and psychological perspective---give us enough reason to revisit VR. We can aggregate such changes in the following categories:

\subsection{Hardware}
\label{sec:hardware}

When it comes to technology, the most remarkable progress was achieved regarding the three following characteristics of VR setups: \textit{affordability}, \textit{comfort}, and \textit{features}. 

\textbf{Affordability.} In general, the price tag is not important in research. However, readily available head-mounted displays (HMDs) created a joint base for VR experiments and increased reproducibility. Thanks to default setups, such as the HTC Vive ecosystem, researchers began to share their testbed implementations and best practices. This interchange enhanced the overall robustness and transferability of the results. 


\textbf{Comfort.} In the past, wearing a heavy and chunky HMD unavoidably introduced bias due to discomfort, which limited the potential of VR. Nowadays, HMDs have an overall weight of around 500 grams and can be considered at least bearable for a prolonged period. More importantly, the show-stopping and dangerous cable clutter finally disappeared, be it in case of all-in-one solutions, such as the Oculus Quest, or even for desktop VR variants, such as the HTC Vive Pro with a wireless adapter.

\textbf{Features.} Affordable and comfortable HMDs compete on the market, and this competition ensures that manufacturers have to add unique features to their products. Hence, users finally receive acceptable display resolutions and an increased---yet still narrow---field of view. Moreover, VR setups now include sophisticated and precise controllers to enhance the interaction with VR content. And, perhaps most importantly, room-scale VR paved its way into production, which allows us to utilize the most realistic locomotion technique---natural walking.

\subsection{Software}

The progress of VR is not only driven by hardware. Advances in computer vision were a crucial step towards enabling room-scale setups, be it in case of desktop-based infrared tracking approaches (e.g., HTC Vive) or standalone optical tracking solutions (e.g., Oculus Quest). On the developer (and researcher) side, the increasing spread of consumer-grade HMDs gave rise to the integration of VR hardware support into popular frameworks and game engines, such as Unity 3D~\cite{unity}. This integration, in turn, opens VR to a broader mass of potential developers, including indie companies. 

More importantly, such a joint base facilitates the exchange of scientific results, testbed scenarios, assets, and tutorials. Hence, it is not surprising that the majority of current VR publications rely on Unity 3D or similar engines due to the significantly reduced turnaround time from idea to prototype. This significantly increased pace compared to the past decades is another critical reason for revisiting that discipline.

\subsection{Humans}

VR is an experience designed for humans by humans. Hence, an important factor we have to consider is the target audience. The digital behavior of people changed at a fast pace in recent years. By becoming more digital, our inhibition threshold regarding VR technologies is now lower than ever. For instance, advanced interaction metaphors, such as gestures, are already known from smartphones. Furthermore, 3D cinemas and similar stereoscopic experiences increased our robustness regarding cybersickness.

These behavioral changes potentially have a high impact on a set of VR research areas. In particular, aspects such as locomotion, interaction, and perception are vital candidates to be affected by the change in our (digital) nature. Hence, the thesis places a particular emphasis on these three areas while leaving out some other entities, such as audio.

\section{Scope and Contextual Boundaries}

The virtual experience is based on several sensations. The general impression is formed by what we see (\textit{vision}), touch (\textit{somatosensation}), hear (\textit{audition}), smell (\textit{olfaction}), and taste (\textit{gustation}). Some of these aspects were affected more by the VR advances mentioned above than others. For this reason---and to maintain focus---the thesis concentrates on vision and somatosensation and does not include research that targets audition, olfaction, and gustation. Of course, this does not mean that these areas are not worth further explorations. Particularly regarding olfaction~\cite{yanagida2004projection} and gustation~\cite{narumi2011augmented}, we surely can expect significant advances soon. In the meantime, we take a closer look at the following research disciplines:

\textbf{Locomotion.} The manuscript provides a high-level overview and classification of VR locomotion techniques, be it stationary or walking-based. Moreover, we introduce and discuss two novel approaches based on virtual body resizing and natural walking: \textit{GulliVR} and \textit{Outstanding}. 

\textbf{Interaction.} In this manuscript, we contribute the following advances to this very diverse research area:
\begin{itemize}[noitemsep]
\item \textit{Haptics -} by introducing a self-transforming controller approach
\item \textit{Inventory systems -} by establishing a VR inventory taxonomy
\item \textit{Gestures -} by designing gestures that an audience can easily understand
\item \textit{I/O abstraction -} by developing a framework for arbitrary, multimodal I/O.
\end{itemize}

\textbf{Perception.} VR setups impact the way how we see the (virtual) world and ourselves. We focus on two examples: First, we take a closer look at the \textit{illusion of virtual body ownership}---a phenomenon that allows us to perceive our virtual avatar as our own body. We extend this illusion to nonhumanoid representations, allowing users to make novel experiences by embodying various creatures. Second, we dive into the area of visualization and examine how the dichoptic presentation of a VR setup can be ``exploited'' for \textit{preattentive highlighting}.


\section{Structure of the Synopsis}

\begin{figure}[htbp]
\centering
\includegraphics[width=1.03\columnwidth]{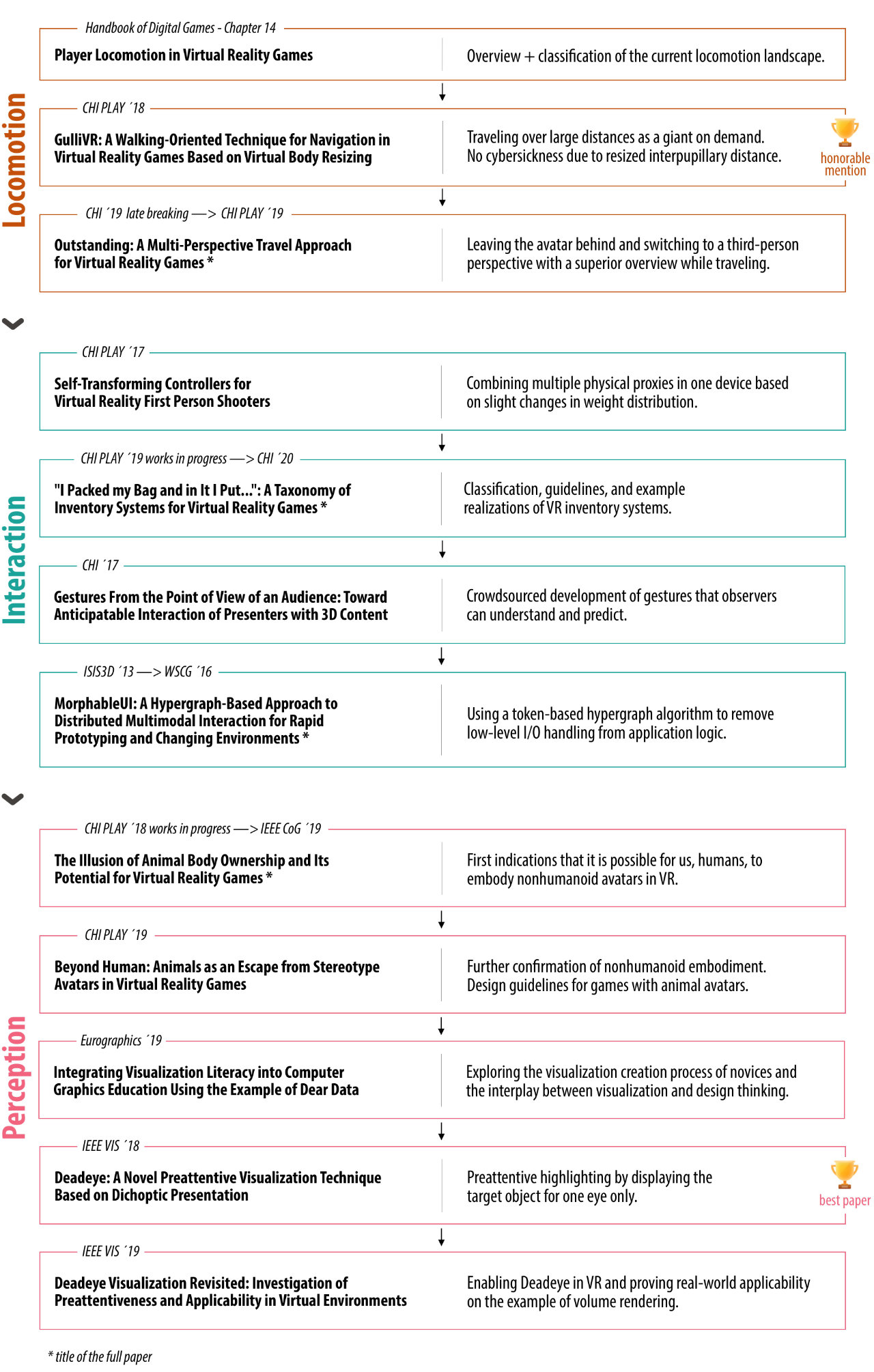}
\caption{The main part of the synopsis and the respective publications at a glance. Note that intermediate manuscripts, such as workshop contributions or works in progress, are not listed as standalone items to maintain clarity.}
\label{fig:synopsis}
\end{figure}

This dissertation consists of two parts: a synopsis and a collection of related publications. The main goal of the synopsis is to establish a red thread through the included papers by grouping and summarizing the conducted research in a high-level manner. For each research topic, the reader obtains a brief motivation, summarized contribution, and an overview of the results. In particular, the insights exposed in the synopsis are not meant to be a standalone report and should be regarded as pointers to the actual manuscripts. For instance, the synopsis does not provide particular $p$-values or source code to prevent a loss of focus.

Overall, the thesis covers multiple research objectives that are thematically organized in three chapters---\textit{Locomotion}, \textit{Interaction}, and \textit{Perception} (cf. \FG{fig:synopsis}). We start by taking a closer look at one of the most crucial aspects of VR, namely locomotion. After establishing a classification of the diverse movement techniques, we discuss two novel approaches for traveling in virtual environments. Both approaches share the common idea of virtual body resizing and eye distance modifications to allow natural walking over large distances without cybersickness.

In the second part, we examine four selected aspects of multimodal interaction. Similar to our walking-inspired contributions in the case of locomotion, we initially focus on the most realistic way to interact with objects in VR. More precisely, we explore the benefits of physical proxies on the example of a self-transforming controller that mimics different virtual objects through a weight-shifting approach. Such interaction with multiple objects requires a useful way to store and represent these entities in VR. For this purpose, we construct a taxonomy for VR inventories and elaborate it on three example scenarios. Then, as a sidestep, we escape user-centered interaction research and consider the spectators instead. Our question in mind is whether and how it is possible to establish gesture-based interaction that is comprehensible and predictable by an audience. Finally, after being confronted by such a diversity of I/O modalities, we see how a hypergraph-based abstraction model can be used to hide this complexity from users and developers to foster the usability of multimodal I/O setups.

The third part of the synopsis is dedicated to perceptional phenomenons that we can achieve thanks to VR. In other words, we focus less on improving VR per se and instead seek for possibilities to benefit from the unique features of such setups. We familiarize ourselves with the so-called illusion of virtual body ownership and transfer such feelings of embodiment onto nonhumanoid avatars. In particular, we see how VR allows us to become an animal and to gain novel experiences in the role of such creatures. Furthermore, we delve into the area of preattentive vision and learn how VR technology can be utilized to highlight objects in visualization tasks.

The synopsis is rounded off by a general discussion about the conducted research. We look critically at the overall contribution of this dissertation and express our final thoughts regarding the possible impact of our results on the status quo of virtual experiences.

\chapter{Locomotion}
\label{sec:locomotion}

Virtual environments evoke our sense of adventure and our urge for exploration. Hence, locomotion ever since had an important role in VR research. Consequently, we dedicate the first chapter to answer the question of how to get from point A to point B in a virtual environment. More precisely, we focus on VR games as a directly affected application area, as it allows us to initiate a more in-depth discussion. Hence, we replace the generic ``user'' by the more fitting term ``player''. Naturally, the presented research is also applicable to other VR domains with no or minor restrictions.

In particular, this chapter outlines three contributions: an overview of the current VR locomotion landscape~\cite{handbook} and two novel approaches that emphasize natural walking as the most natural way of movement. Accordingly, we begin by drawing a big picture of locomotion techniques to provide orientation and to establish a foothold on basic concepts, such as presence and cybersickness. Hereafter, we discuss the idea of virtual body resizing that lays the foundations for the two novel techniques \textit{GulliVR}~\cite{gullivr} and \textit{Outstanding}~\cite{outstanding1,outstanding2}.

\section{VR Basics: Presence and Cybersickness}

Before delving into locomotion and VR research in general, we need to revisit certain core mechanics of such setups. In addition to providing stereo images, VR headsets also track the head orientation, which allows us to look around in VR like in the real world. This impression is often described as a feeling of ``being there''~\cite{heeter1992being,lombard1997heart}. Other popular wordings include the two terms \textit{presence} and \textit{immersion}---sometimes in an interchangeable manner. Throughout the dissertation, we utilize immersion~\cite{cairns2014immersion} when focusing on the technical quality of VR hardware ~\cite{Biocca:1995:IVR:207922.207926, sherman2002understanding}. In contrast, we use presence to describe how immersive setups affect our perception~\cite{slater2003note}. Researchers often target the increase of presence when coming up with novel VR approaches, and VR locomotion is a prominent advocate for such presence-enhancing research~\cite{slater1995taking}. The most common methods to measure presence include the Presence Questionnaire (PQ)~\cite{Witmer.2005} and the Igroup Presence Questionnaire (IPQ)~\cite{Schubert.2003, Schubert.2018}. Also, the Immersive Tendencies Questionnaire (ITQ)~\cite{witmer1998measuring} can be administered before the actual survey to check participants' tendencies to get immersed in an activity or fiction. 

The feeling of presence imposes high demands on VR software and hardware. Minor faults or technical issues, such as a slightly offset locomotion or a sudden drop in frame rate, can have severe consequences regarding the players' well-being and result in \textit{cybersickness}~\cite{stanney1997cybersickness,laviola2000discussion}. Other prominent terms often used in this context are \textit{simulator sickness}~\cite{kolasinski1995simulator} and \textit{motion sickness}~\cite{money1970motion,hettinger1992visually,ohyama2007autonomic}. In contrast to the manifold reasons of cybersickness, the main cause of simulator sickness is an incorrectly adjusted simulator~\cite{kennedy1989simulator}, i.e., a rather technical problem.

Cybersickness involves symptoms such as nausea, eye strain, and headaches. The main reason behind that negative phenomenon is a mismatch in our vestibulo-ocular system: Our vestibular system senses acceleration that ideally matches the visual input. When these signals do not match, the symptoms mentioned above are likely to occur~\cite{reason1975motion}. Three popular explanations~\cite{laviola2000discussion} exist for such an undesired body reaction: poison theory, postural instability theory, and---the most prominent---conflict theory. Hettinger et al.~\cite{hettingerVection} also mentioned vection as a possible reason behind cybersickness. Vection describes the feeling of movement that relies only on our visual system and occurs when we, e.g., sit in a standing train and observe another train that is currently accelerating. Vection is influenced by several factors, including the field of view (FOV), the alignment and proximity of moving objects, and the optical flow rate. For instance, the combination of a large FOV and fast-moving objects that account for a large proportion of players' view amplify the perceived vection and smooth the way to cybersickness. Hence, limiting the FOV is one of the possible approaches to reduce cybersickness~\cite{fernandes2016combating,lin2002effects}.

Regarding VR locomotion, we recommend keeping cybersickness in mind and avoiding cognitive mismatches where possible. However, this guideline should not be pursued at all costs: ``sickness-save'', stationary scenes without any locomotion will potentially miss out on a wide range of benefits that VR has to offer. Moreover, as shown by von Mammen et al.~\cite{von2016cyber}, games might even benefit from an artificially induced cybersickness in some instances.

\section{Overview of the VR Locomotion Landscape}

\begin{figure}[t!]
\centering
\includegraphics[width=1.0\columnwidth]{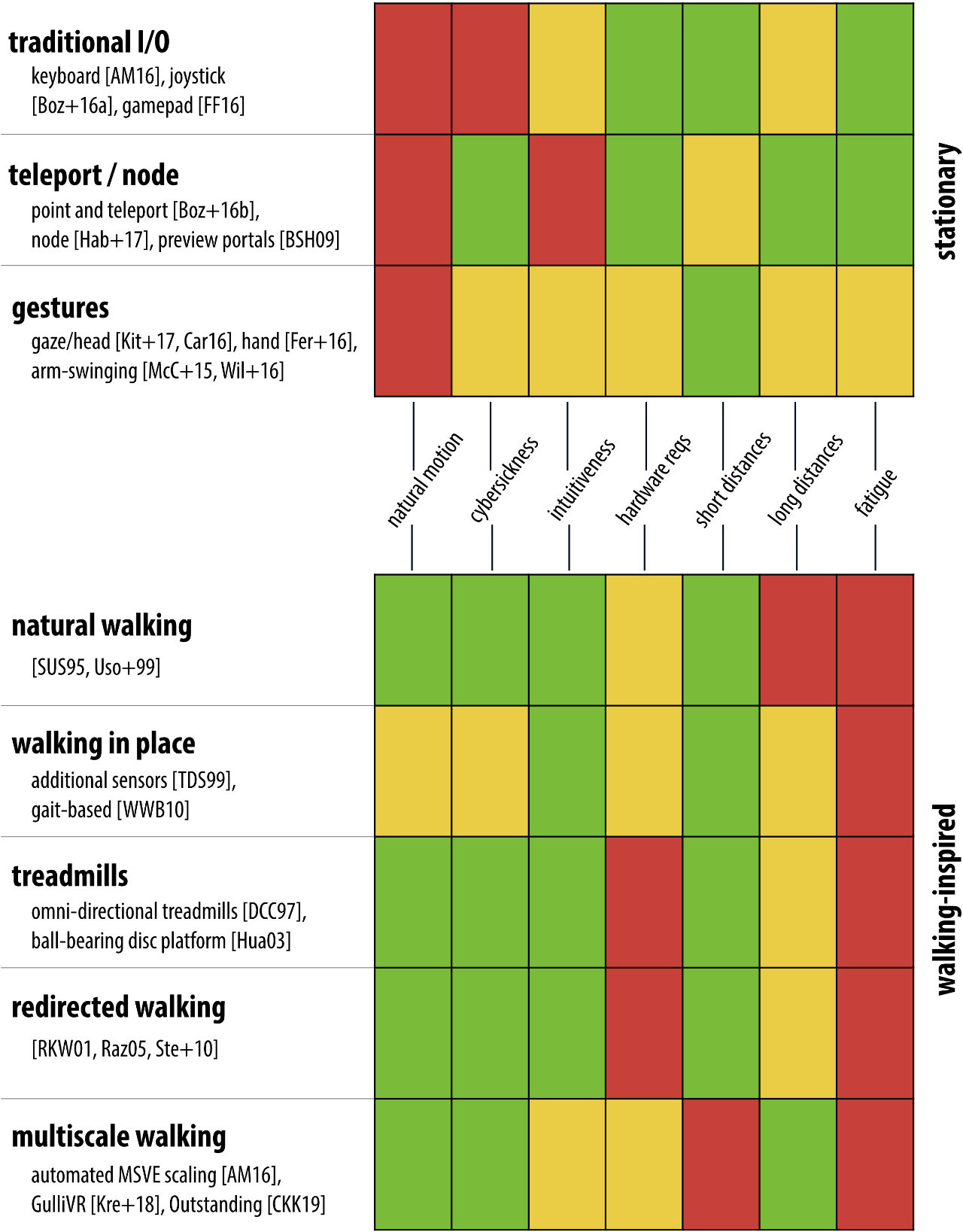}
\caption{An overview of VR locomotion techniques including their benefits and drawbacks. Green color means that this attribute is a known advantage of the technique, yellow stands for limited value, and red is rather a disadvantage.}
\label{fig:locomotion}
\end{figure}

In recent years, research~\cite{boletsis2017new} and the game industry~\cite{Habgood:2017:HLP:3130859.3131437} created a plethora of varying locomotion approaches. As often the case, there is no ``right'' answer which one to pick for an upcoming VR game. Hence, the main purpose of our research article on player locomotion~\cite{handbook} was to analyze the broad spectrum of possibilities to move in VR and to serve as a starting ground for further research and (game) development. The summarized approaches range from stationary methods that rely on gamepads to advanced redirected walking techniques that trick our perception into enabling unrestricted natural walking in a limited space. 

\FG{fig:locomotion} outlines the proposed division in stationary and walking-oriented approaches. Stationary techniques, such as gamepad or gesture-based controls, do not involve a physical movement of the player. Hence, the main reasons for choosing such approaches are the lower hardware requirements, as we do not have to track the absolute player position, and the reduced fatigue, as players can remain seated during a gaming session. On the downside, standing still while moving in VR usually comes with the risk of cybersickness due to the cognitive mismatch. Therefore, previous research~\cite{medeiros2016effects,yao2014oculus} suggests short, fast movements with no acceleration as one measure to combat cybersickness in such cases.

In contrast to stationary techniques, most of the approaches inspired by natural walking are intuitive and robust concerning cybersickness. A significant amount of research has confirmed the resulting superior realism of physical walking in VR and its positive effect on the perceived presence~\cite{slater1995taking,usoh1999walking,ruddle2009benefits,waller2013sensory,ruddle2006efficient}. Unfortunately, unrestricted natural walking is hardly achievable, as the physical room size imposes an insuperable obstacle. Hence, approaches falling in the walking-inspired category attempt to overcome this limitation in various ways---be it physical treadmills or redirected walking.

Apart from the outlined classification, our work contains a set of design implications to facilitate the decision-making process on locomotion in research and (game) development. We base our suggestions on the following three questions:
\begin{itemize}[noitemsep]
\item \textit{What is our target audience?}
\item \textit{How important is exploration?}
\item \textit{What else happens during locomotion?}
\end{itemize}
A general response to these questions can be extracted from \FG{fig:locomotion}. For a more detailed discussion of each aspect, we invite the reader to revisit the full publication where we provide the corresponding design considerations.

\nocite{argelaguet2016giant}
\nocite{bozgeyikli2016locomotion}
\nocite{fernandes2016combating}

\nocite{bozgeyikli2016point}
\nocite{bruder2009arch}
\nocite{Habgood:2017:HLP:3130859.3131437}

\nocite{kitson2017comparing}
\nocite{cardoso2016comparison}

\nocite{harris2014human}

\nocite{ferracani2016locomotion}
\nocite{piumsomboon2013user}

\nocite{wilson2016vr}
\nocite{mccullough2015myo}

\nocite{slater1995taking}
\nocite{usoh1999walking}

\nocite{templeman1999virtual}
\nocite{wendt2010gud}

\nocite{Darken:1997:OTL:263407.263550}
\nocite{huang2003omnidirectional}

\nocite{razzaque2001redirected}
\nocite{razzaque2005redirected}
\nocite{steinicke2010}

\nocite{argelaguet2016giant}
\nocite{Krekhov:2018:GVRA}
\nocite{cmentowski2019outstanding}

\section{Novel Approaches: GulliVR and Outstanding}

Our research contributes to the field of walking-inspired techniques. We believe that the significant advantages of natural walking, such as the mentioned increase in presence~\cite{slater1995taking,usoh1999walking,ruddle2009benefits,waller2013sensory,ruddle2006efficient} or the improvement of our cognitive map~\cite{ruddle2011walking}, are worth our research efforts to deal with the fundamental problem of physical space limitations. 

In the broader sense, our approaches can be classified as a multiscale virtual environment navigation~\cite{lambooij2009visual,argelaguet2016giant,Zhang:2002:SIM:571878.571884,kopper2006design}. Common locomotion methods are based on a 1:1 ratio between the physical and the real world, i.e., the traveled distance in the physical room roughly corresponds our virtual footprint. In contrast, we fit the virtual environment entirely into our living room by rescaling the player. In the remainder of this chapter, we briefly outline the core idea of virtual body resizing and how it manifests in our two locomotion approaches. Finally, we share some ideas about the general applicability of our work and its overall impact on the VR locomotion landscape.

\subsection{Virtual Body Resizing}

Our core idea is to enlarge the virtual body of the player on demand. Such a transformation allows the player to travel vast distances in a room-scale environment using natural walking. We can obtain the same results by shrinking the size of the world instead of enlarging the player. However, we would not recommend that approach in practice for performance reasons. For instance, having a fully resizable environment usually interferes with baked lighting and also favors floating point precision errors.

\begin{figure}[t]
\centering
\includegraphics[width=1.0\columnwidth]{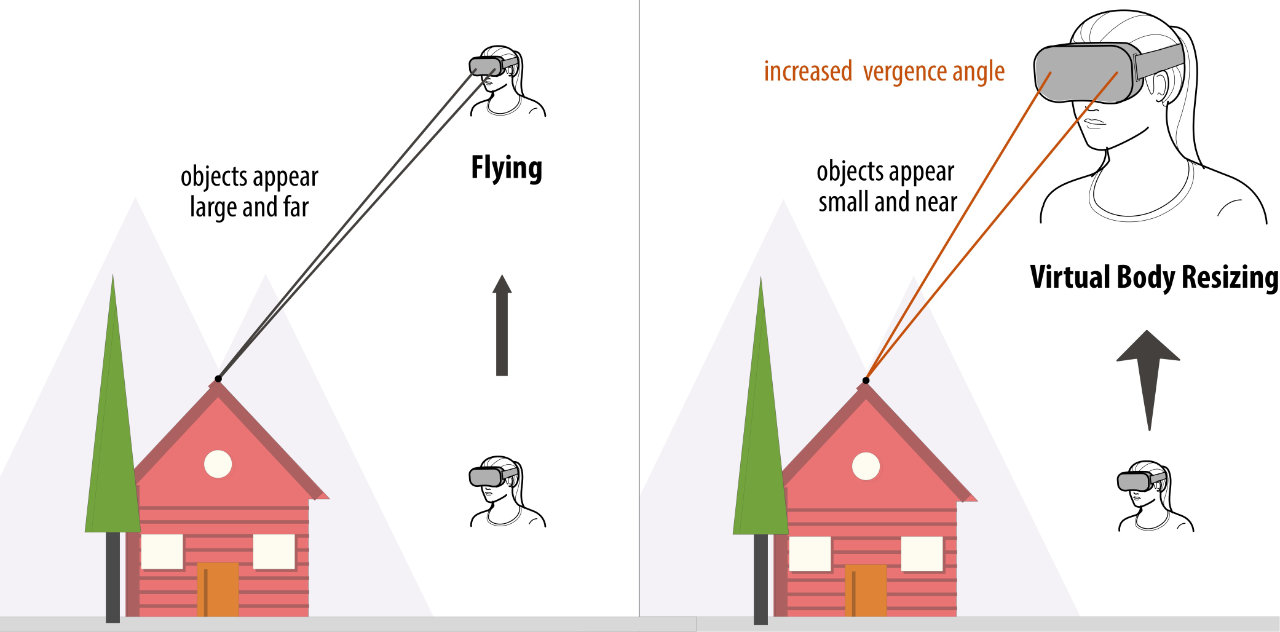}
\caption{The increased eye distance results in a larger vergence angle, altering the size/distance perception of objects and evoking the feeling of being a giant. Physical movement is perfectly aligned with visual feedback, which obviates cybersickness.}
\label{fig:resizing}
\end{figure}

At first glance, the idea is similar to a common flying approach, as both variants share the same camera position and velocity. However, flying is known for its severe cybersickness due to the cognitive mismatch of physical and virtual movement speeds. In contrast, enlarging the player comes with an increase of the interpupillary distance, as depicted in \FG{fig:resizing}. Although tiny variations of the interpupillary distance have been shown to have no measurable impact on size judgments~\cite{best1996perceptual} and can even be applied unnoticed~\cite{ware1998dynamic}, setting the modeled eye separation to a significantly different value compared to the physical eye distance results in so-called \textit{false eye separation}~\cite{cho2014evaluating}. The resulting perceived image causes an altered size perception compared to real objects~\cite{wartell1999analytic}, leading to a miniature world perspective. Renner et al.~\cite{renner2015influence} and van der Hoort et al.~\cite{van2011being} also reported similar findings, confirming that increasing the stereo base (i.e., the modeled eye distance) makes objects appear nearer and smaller. This altered perception allows us to minimize the risk of cybersickness, as players feel like they are walking as giants (no mismatch) and not like they are artificially floating or flying.

\subsection{GulliVR: Walking as a Giant}

The \textit{GulliVR} navigation metaphor~\cite{gullivr} utilizes virtual body resizing to transform players into giants on demand. As giants, players can travel over large distances within a few steps, and switch to normal size once the destination is reached (cf. \FG{fig:gullivr}). Although the underlying mechanism is straightforward,  we have to consider several degrees of freedom for proper functioning. First and foremost, we need to decide how fast the transition between the two scales should be. By means of a pretest, we found a sweet spot at $t = 0.005 \cdot \textrm{Scale}_\mathit{giant}$. As an example, our testbed scenario utilized $\times100$ as a scaling factor, resulting in a transition of half a second. Slower transition times bear the risk of cybersickness, while faster or instant transitions favor player disorientation.

\begin{figure}[t!]
\centering
\includegraphics[width=1.0\columnwidth]{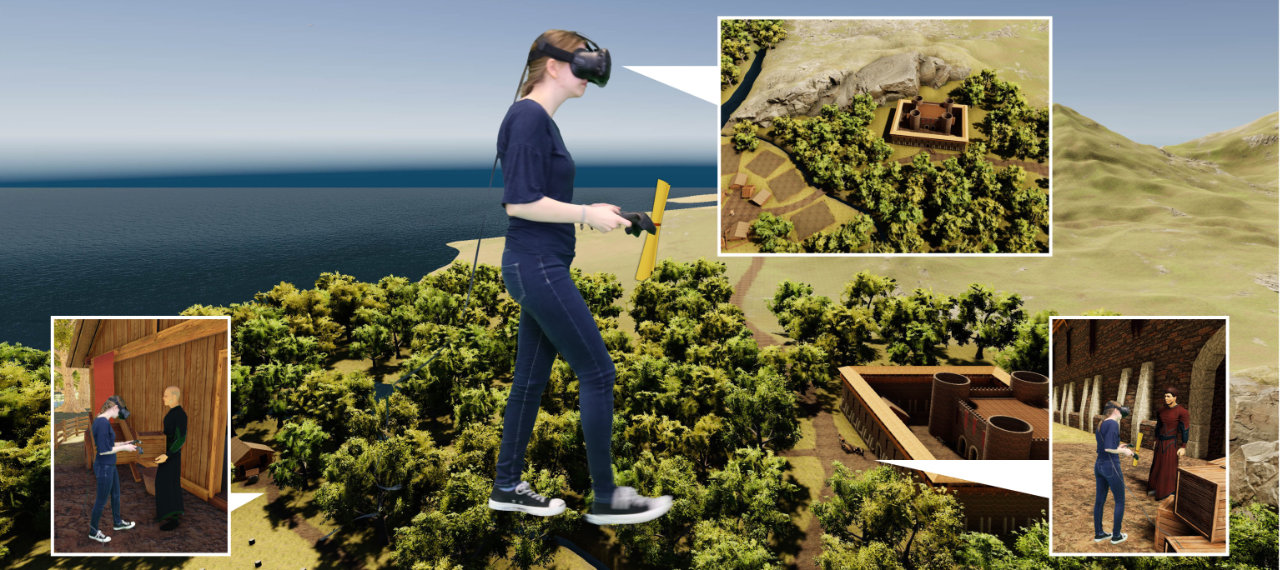}
\caption{\textit{GulliVR:} players turn into giants and traverse large distances without cybersickness due to the adjusted interpupillary distance.}
\label{fig:gullivr}
\end{figure}

The main reason for us to emphasize natural walking was the expected increase in presence. Hence, we conducted a user study that compared \textit{GulliVR} with the established point and teleport locomotion approach~\cite{bozgeyikli2016point}. Our results confirmed our hypothesis regarding presence. Furthermore, players in the \textit{GulliVR} group walked significantly more and did not show any signs of cybersickness.

In addition to these fundamental advantages, our work established different possibilities for leaving the giant mode. In particular, we discussed the unguided, vanilla version, a crosshairs-based extension, and an automated pulling system, as depicted in \FG{fig:aiming}. The latter option is what we utilized in our testbed game: leaving the giant mode pulled the player discreetly toward the nearest point of interest to avoid unnecessary readjustments and frustration.

To summarize, \textit{GulliVR} successfully introduced the core approach of giant-like traveling in VR games. The idea was later picked up by Abtahi et al.~\cite{abtahi2019m} and Cmentowski et al.~\cite{outstanding1,outstanding2}. In particular, these works confirm our finding that the increase of the interpupillary distance in giant mode prevents cybersickness, which is an essential requirement for the applicability of such locomotion methods.

\begin{figure}[t!]
\centering
\includegraphics[width=1.0\columnwidth]{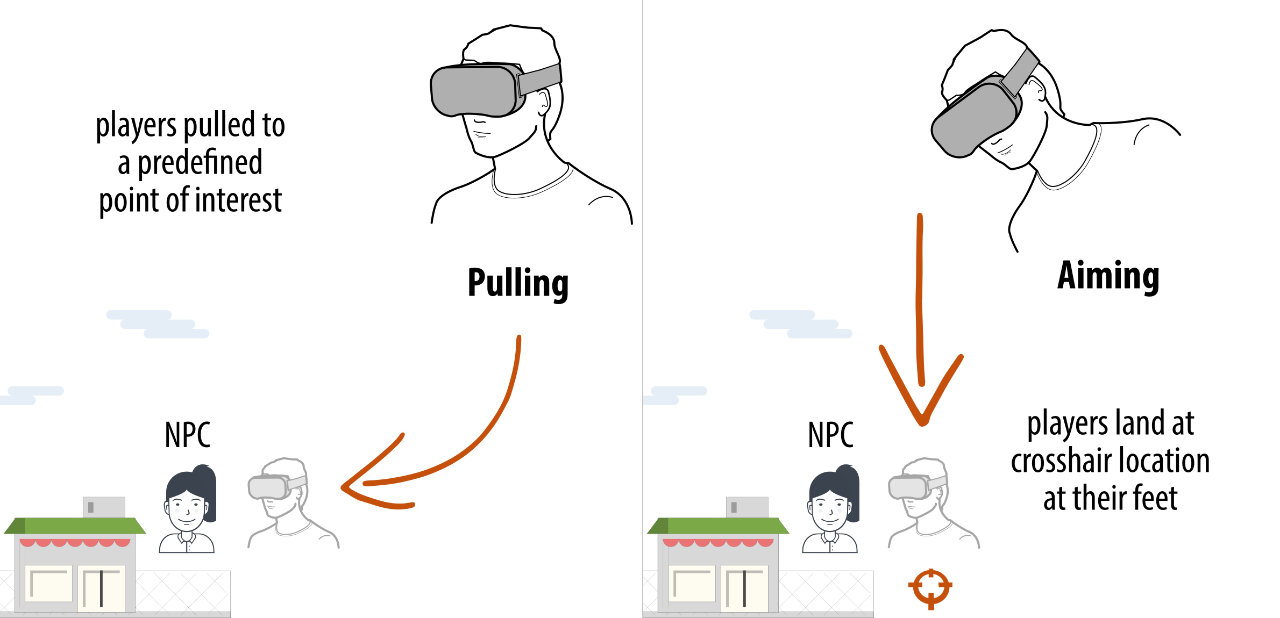}
\caption{Two extensions of GulliVR to enable a precise transition from giant mode back to normal mode. Pulling adds a horizontal translation toward the nearby point of interest, whereas aiming displays a crosshair to indicate the destination location.}
\label{fig:aiming}
\end{figure}

\subsection{Outstanding: Leaving the Avatar Behind}

In contrast to \textit{GulliVR}, which relies on a consistent first-person view, the \textit{Outstanding} technique~\cite{outstanding1,outstanding2} connects the virtual body resizing to a dynamic switching between first-person and third-person perspectives. In the unscaled state, players get the typical first-person view to perform short-range exploration by physical walking, detailed observation of local points-of-interest, and basic interactions, such as picking up objects. For large-distance traveling, we once again perform the virtual body resizing. This time, however, the avatar is left behind, and players obtain a third-person bird's eye perspective. We display the virtual avatar at the players' feet to symbolize their original first-person position in the world. This view allows the disembodied players to observe the surrounding area from an elevated view and to command their avatar by setting navigation targets using raycast aiming (cf. \FG{fig:outstanding}).

\begin{figure}[t!]
\centering
\includegraphics[width=1.0\columnwidth]{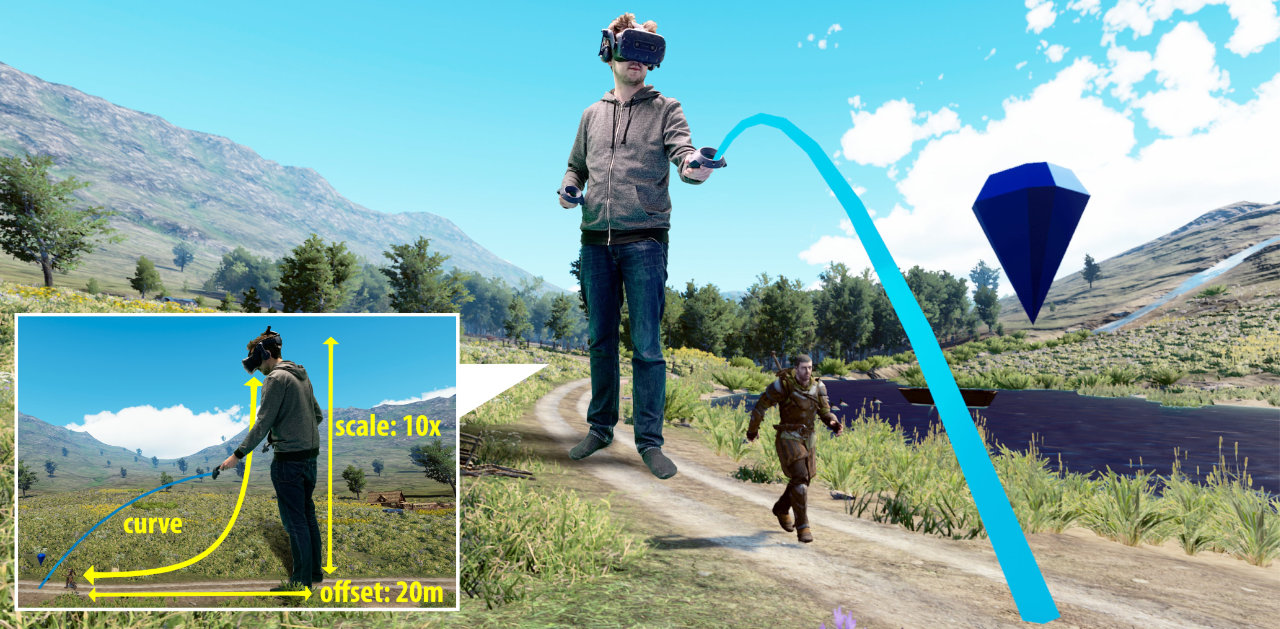}
\caption{\textit{Outstanding:} players can switch to a bird's eye third-person perspective and control their avatar via raycast aiming. The picture-in-picture shows the optimal transition parameters to convey the feeling of embodying or disembodying the avatar.}
\label{fig:outstanding}
\end{figure}

Each perspective has benefits and drawbacks~\cite{salamin2006benefits,gorisse2017first}, i.e., the third-person mode is excellent for environmental perception while the first-person view is superior regarding interaction-intensive tasks. The introduced dynamic perspective switching combines the strengths of both views and achieves uncomplicated, overview-enhancing traveling with interactive, local exploration on demand. 

Similar to \textit{GulliVR}, the transition between the two modes---and even perspectives in this case---required additional iterations and prestudies. In our work, we proposed a fast, dolly-shot-like animation to improve the impression of embodying or disembodying the avatar. We also added a translation backward to achieve a comfortable 45\textdegree~viewing angle after switching to the third-person mode and included a curved animation between both states. This curve emphasizes horizontal disembodiment followed by a steeper vertical growth, as shown in \FG{fig:outstanding}.

To validate our technique, we compared \textit{Outstanding} to the point and teleport locomotion approach. The results of our study confirmed a significant increase in spatial orientation while maintaining high levels of presence, competence, and enjoyment. Our experiments showed that players generally liked the idea of the dynamic switching between different perspectives and that they were able to use the technique without significant problems.  Additionally, the work summarized the critical insights into a set of comprehensive design guidelines and established technical extensions, such as a catching-up mechanism to skip ahead and close the gap between the player and the avatar without switching perspectives.

\section{Concluding Thoughts on Resized Walking}

Apart from a classification of the current locomotion landscape, this chapter introduced two novel locomotion techniques that rely on virtual body resizing. The core idea was the enlargement of players on demand to conquer large distances within a few steps. Our main motivation was to emphasize natural walking as the most intuitive locomotion method, and indeed, according to our results, we succeeded on multiple fronts: 
\begin{itemize}[noitemsep]
\item the adjustment of the interpupillary distance \textbf{prevented cybersickness}
\item the view from above \textbf{increased the spatial orientation}
\item the approaches had a \textbf{positive impact on presence}
\item \textbf{players walked more}, enjoying the advantages of physical movement~\cite{ruddle2009benefits}.
\end{itemize}

However, such contributions always require critical analysis. Neither \textit{Outstanding} nor \textit{GulliVR} remove the room-scale limitation. Even if we can travel much larger distances thanks to rescaling, we will eventually end up in front of a physical wall. For instance, in the case of \textit{GulliVR}, players have to go back to normal scale, walk away from the wall, and re-enter the giant mode to continue traveling. For this reason, we position our approaches as additional locomotion methods and not as standalone solutions. More precisely, we recommend a combination with stationary approaches, such as teleportation, to facilitate such resetting and foster short-distance exploration.

As player resizing comes with specific strengths and weaknesses, we have to embed such techniques carefully in the underlying scenarios. In our testbed scenarios, players encountered a medieval, fantasy-inspired setting, where being a giant potentially fits the narrative. In contrast, we discourage the application of such methods in games with dominant indoor scenery: narrow, multilevel closed environments diminish the advantages mentioned above, such as the increase of spatial orientation, and further complicate targeting when switching back to a regular scale.

In addition to such scenery-related decisions, we recommend to consider the pace of the game: dynamic first-person shooters are less likely to benefit from such exploration-oriented techniques than slow-paced role-playing games or 3D adventures relying on big open worlds. However, note that players could easily get spoiled from seeing too far ahead, which requires additional techniques like the fog of war.

An important lesson that we learned from our evaluations is that players desire additional interactions while traveling as giants. For instance, in the case of \textit{Outstanding}, players have to wait for their avatar to reach the destination, and just looking around quickly gets annoying. Digital games keep players engaged by introducing new events regularly. Similarly, we propose to include additional incidents that force players to switch back to regular size or to add specific actions as a giant. The latter option was highly popular in the \textit{GulliVR} study, and players enjoyed their superhuman abilities while picking up and moving large objects.

To summarize, our research has shown how virtual body resizing can be used to design locomotion techniques with unique advantages, such as increased spatial orientation and fast travel without cybersickness. We suggest that rethinking player movement from scratch, as we did with \textit{Outstanding} and \textit{GulliVR}, is a more promising way rather than porting locomotion 1:1 from non-VR games. ``VR-first'' approaches allow us to take full advantage of the benefits that VR setups have to offer, and we believe that our and similar research efforts can pave the way to even more engaging and fascinating player experiences and game mechanics in VR.

\chapter{Interaction}
\label{sec:interaction}


In the previous chapter, we mainly dealt with the issue of getting from point A to point B. However, locomotion is just a basic need in VR. To create an engaging user experience and fully benefit from immersive setups, we need intuitive, user-tailored interaction solutions that amplify our feeling of being in the virtual world.

This chapter covers four subareas of interaction and explores ways to enhance the scope of user actions in VR. Throughout our related research, our primary motivation was to make user interaction in VR as realistic and intuitive as possible. In particular, we achieved these goals by developing physical proxies~\cite{controller}, rethinking virtual item representations~\cite{inventories1,inventories2}, designing comprehensive gestures~\cite{gestures}, and providing a meaningful I/O abstraction layer~\cite{muiold,mui} to bring all these components together into a coherent VR ecosystem.

We start with object manipulation in VR and the possibilities to realize haptic feedback for such interactions. More precisely, the first section focuses on the design of input controllers that can accommodate multiple virtual objects via shapeshifting.

\section{Haptics: A Self-Transforming VR Controller}

One easy way to break presence in VR is to touch a virtual object without getting a proper tactile feedback~\cite{Simeone:2015:SRU:2702123.2702389}. Hence, input controllers play an essential role in our virtual experience~\cite{jennett2008measuring,young2016usability,birk}. In this section, we consider the design of such controllers on the example of first-person shooters~\cite{zaranek2014performance}. We may assume that a well-crafted controller that feels like its virtual counterpart~\cite{brown2015beyond}, i.e., a gun, can increase the feeling of presence. For instance, the Vive controller resembles a gun handle with a trigger, while other devices mimic the shape of two-handed guns to provide players more realistic haptics for games with such weapon types. Note that the actual appeal of such devices is not relevant, as players do not see the controller while being in VR.

\begin{figure}[t]
\centering
\includegraphics[width=1.0\columnwidth]{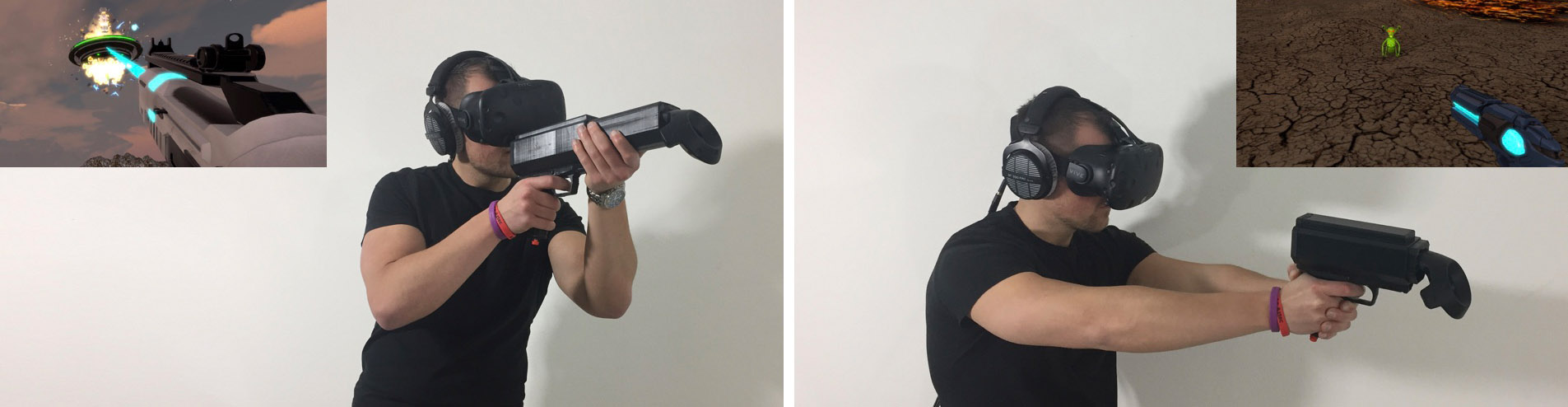}
\caption{Our self-transforming controller simulating a two-handed laser rifle (left) and a one-handed blaster (right). The transformation is triggered by a button at the bottom of the handle and works based on a motor and two telescopic tubes. A built-in Vive controller is used for tracking.}
\label{fig:controller}
\end{figure}

Unfortunately, such devices can not reflect the in-game weapon switching, which is a crucial element of most shooter games. Relying on a one-handed, pistol-like controller for a game where the current weapon is a two-handed rifle is less realistic than a two-handed input device, and vice versa. For this issue, our research proposed a self-transforming controller~\cite{controller}, as depicted in \FG{fig:controller}. Our controller adjusts its shape and handling according to its game representation, i.e., the device feels and behaves similarly to a pistol in the first state, and similarly to a rifle in the second state. The transformation is done via a telescopic, motor-driven mechanism.

Our research contributes to the body of literature related to shapeshifting input devices~\cite{shape1,shape2,shape3,shape4,shape5,shape6}. Similar to Zenner et al.~\cite{zenner}, we rely on weight shifts in the device to modify the rotational resistance and the perceived inertness. Our design thinking workshop and subsequent eveluations revealed the following insights:
\begin{itemize}[noitemsep]
\item controllers should be \textbf{lighter} than real guns (< 0.8 kg)
\item controllers should be \textbf{shorter} than real guns (guns: < 25 cm, rifles: <40 cm)
\item \textbf{subtle weight shifts suffice} (in our case, a 1:1.7 ratio of gun to rifle).
\end{itemize}

Our results confirmed that such an adaptive approach outperforms default VR controllers concerning scales such as appearance, efficiency, authenticity, experienced realism, and flow. Most importantly, 96 \% of the participants reported liking the overall concept of a self-transforming controller. 

Based on this positive feedback, we suggest applying the same design approach to other weapons or virtual objects. For instance, the inclusion of a second rack that transports an equal weight toward the player allows simulating shoulder-fired missile weapons, which allows us to cover most of the weapon classes for first-person shooters. This kind of multifunctional physical proxies is also beneficial for other game genres, such as role-playing games or 3D adventures. In this regard, a retractable handle would enhance the design space with sword-shaped virtual objects, and simulate a switch between, e.g., a short knife and a two-handed sword.

\section{Inventory Systems: Organizing Objects in VR}

\begin{figure}[t]
\centering
\includegraphics[width=1.0\columnwidth]{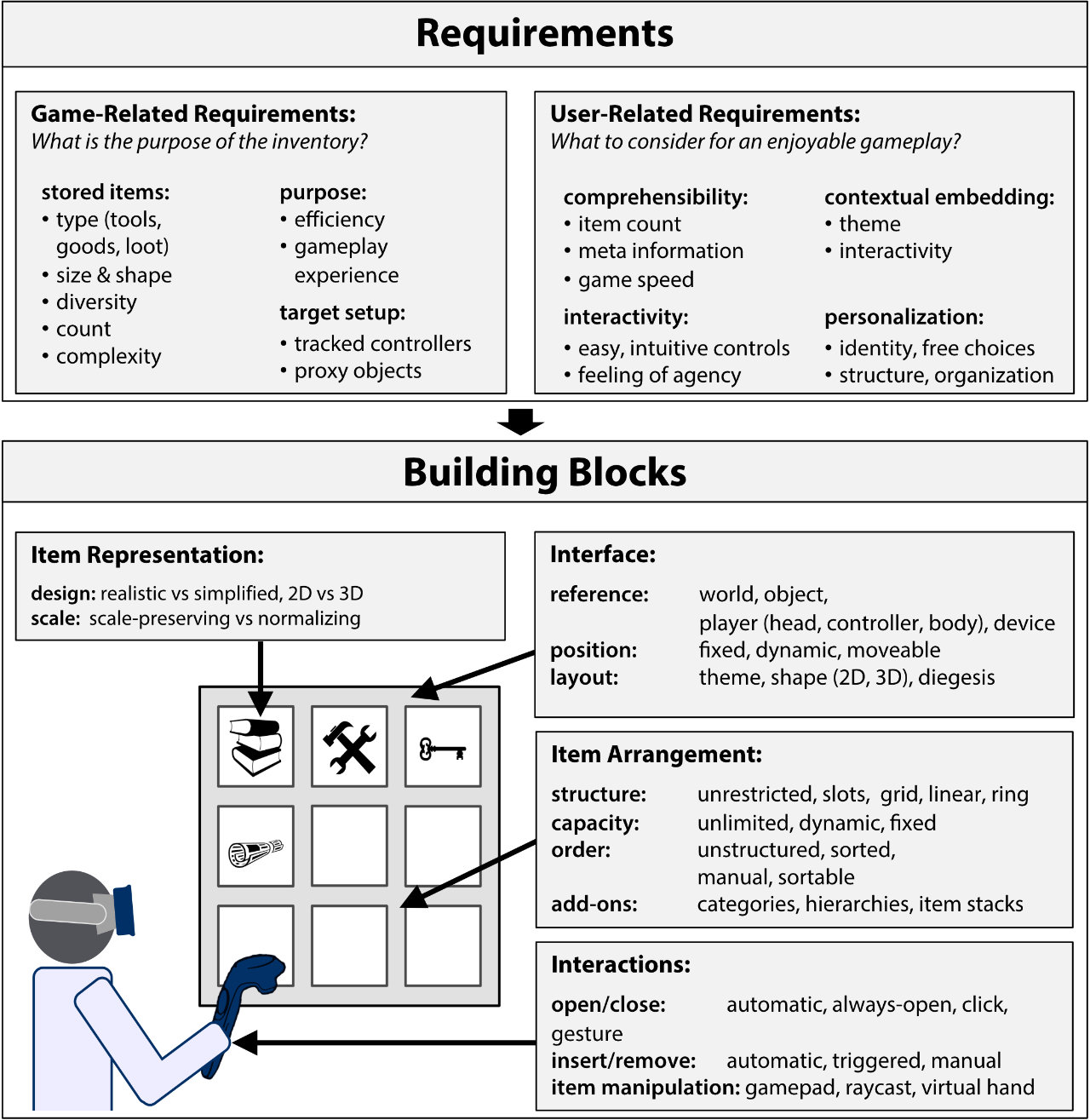}
\caption{Our taxonomy of inventory systems in virtual environments. This figure is read from top to bottom, starting with the requirements that we should take into account before and while designing inventories. The considerations are used to select design choices for each building block at the bottom.}
\label{fig:inventories}
\end{figure}

Transformable controllers that represent multiple virtual objects are a significant advancement toward a more realistic interaction in VR applications and games in particular. Being equipped with multiple virtual objects requires efficient ways to manage such items. Storage interfaces, best known as \textit{inventories}~\cite{wegner2017comparison}, are among the most commonly used features in nearly every game genre. Although carrying multiple items appears like a natural addition to interaction-oriented gameplay, most VR developers refrain from using inventories at all. For this reason, the goal of our research~\cite{inventories1,inventories2} was to understand the peculiarities of VR inventory design. We applied the following scientific methods to achieve this objective:
\begin{itemize}[noitemsep]
\item \textbf{in-depth developer interviews} to determine real-world challenges and pain points related to inventory design in VR 
\item \textbf{literature reviews} to sublime best practices on VR menus in general
\item \textbf{games analysis} via grounded theory approach~\cite{charmaz2007grounded, glaser1968discovery} to identify the essential building blocks and characteristics of VR inventories.
\end{itemize}

The resulting taxonomy is depicted in \FG{fig:inventories}. Also, our publication exposes a set of design implications and demonstrates the practical use of our taxonomy in action. More precisely, the work introduces and evaluates three manifestations of VR inventories: a flat grid, virtual drawers, and a magnetic surface (cf. \FG{fig:inv_examples}).

\begin{figure}[t]
\centering
\includegraphics[width=1.0\columnwidth]{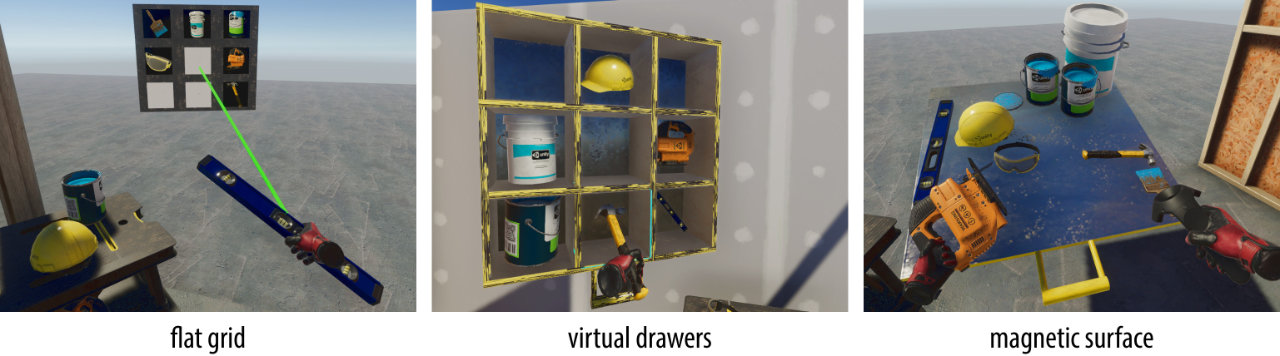}
\caption{Three example realizations based on our taxonomy. \textbf{Flat grid:} a 2D overlay and virtual raycasts are used to achieve fast and straightforward item management. \textbf{Virtual drawers:} items that are inserted using physical actions and scaled to equal sizes. \textbf{Magnetic surface:}
free item placement allows for a precise positioning; items stick to the surface until removed.}
\label{fig:inv_examples}
\end{figure}

The outlined work contributes to the area of VR interaction by decomposing the inventory design process into requirements and building blocks. The resulting taxonomy can be used to facilitate and improve the decision-making of researchers and game developers. There is a need for further evaluations of particular inventories. We assume that such additional studies will allow our community to create a big picture of the interplay between the individual building blocks. This understanding, in turn, will further improve the status quo of user interaction in virtual environments.

\section{3D Gestures: Interaction for the Audience}

So far, we have seen how to realize direct object manipulation with transformable physical proxies and how to manage virtual objects in virtual inventories. As usually the case in human-computer interaction, we explicitly focused on the user. This section breaks with this tradition and focuses on the observers, i.e., the audience instead. More precisely, the outlined work~\cite{gestures} explores whether and how gesture-based interaction can be designed to be understandable by someone who stands and watches aside.

Why is it important to understand our interaction? Firstly, acting or playing in VR is often an experience that we can not easily share due to hardware requirements. In this case, our audience can benefit from clear, unambiguous gestures, which make it easy to follow and understand our actions~\cite{Hespanhol:2012:IIE:2307798.2307804,Grandhi:2011:UNI:1978942.1979061}. Secondly, audience-friendly and anticipatable gestures can be helpful during presentations~\cite{7472190,Curtis:2016:SIA:2993148.2993194,Cuccurullo:2012:GAP:2254556.2254584,Tan:2010:GAI:1873951.1874041}, as they facilitate the process of message transportation~\cite{importance1, importance2, importance3}.

\begin{figure}[t]
\centering
\includegraphics[width=1.0\columnwidth]{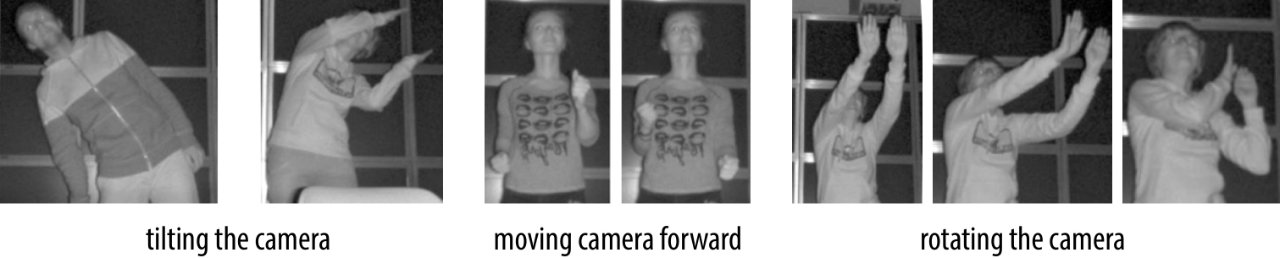}
\caption{In the Wizard of Oz experiment, participants were asked to perform body movements and complete a set of tasks, such as moving (the camera) forward. The wizard monitored the participants via the Kinect camera and triggered the corresponding behavior within the target application.}
\label{fig:wizard}
\end{figure}

We relied on a Wizard of Oz experiment~\cite{Althoff2004, Dahlback:1993, Maulsby:1993, Salber:1993} with representatives from a typical audience to design multiple sets of anticipatable gestures (cf. \FG{fig:wizard}). By filtering and clustering the results, we established the following three distinct sets: bimanual gestures, one-armed gestures, and full-body movements, as shown in \FG{fig:gestures}.

\begin{figure}[b]
\centering
\includegraphics[width=1.0\columnwidth]{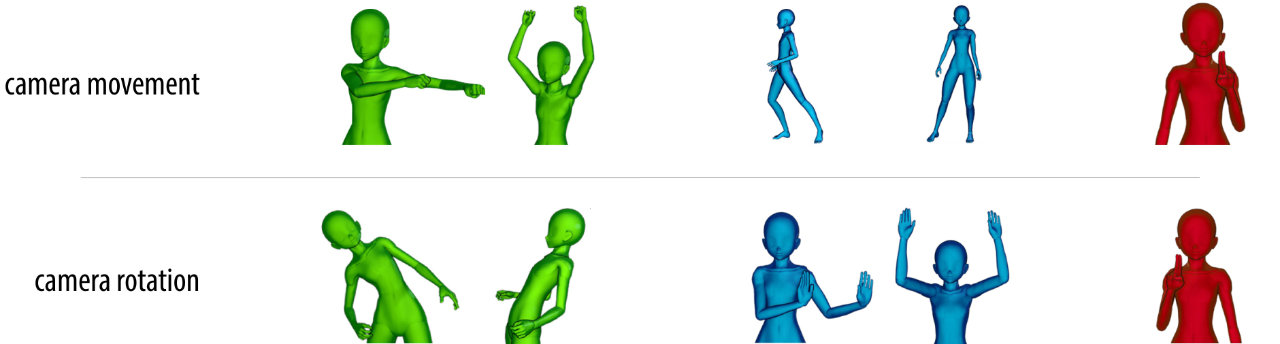}
\caption{Gestures for two example tasks. The green set mostly contains bimanual gestures. The blue set adds weight transfer. The red set consists of minimalist, one-armed gestures, and is the only set that was not predictable by the audience in our study.}
\label{fig:gestures}
\end{figure}


An online evaluation revealed that two of the three gesture sets were indeed anticipatable, which supports our suggested approach of gesture crowdsourcing and emphasizes the benefits of gestures for comprehensible 3D presentations. In particular, we determined that gestures that rely on weight shifting perform very robust in terms of possible misinterpretations. Ambiguous configurations, i.e., where different arms lead to different interactions despite executing the same gesture, are rather hard to grasp and should be avoided.

For future research and applications, we recommend the following guidelines for anticipatable gestures:
\begin{itemize}[noitemsep]
\item keep gestures as \textbf{distinct} as possible
\item use \textbf{full body movements} (e.g., weight shift) rather than minimalistic gestures
\item provide \textbf{analog behavior} for both arms.
\end{itemize}
Such gestures depend on the target application and the environment (e.g., the given VR setup). In our case, we considered the example of a digital planetarium. Hence, the gestures might need an adaptation for a different use-case. Furthermore, we suggest to examine the learning curve of an audience in a future study, i.e., the aspect of how easily a gesture can be learned might play an even more important role than the initial anticipation. We also encourage similar evaluations for VR games, especially in combination with live streaming video platforms where player actions in VR need to be understood by thousands of viewers.

\section{MorphableUI: I/O Handling in VR Applications}

Sophisticated gestures and (transformable) controllers are only a few examples of interaction modalities that are possible in VR. Speech input, mobile devices, or gaze-based controls---we have a plethora of contemporary I/O modalities and devices at our disposal. This freedom comes at a cost, as the engineering workload involved in making applications fully adaptable in this sense is very high. Hence, we need a more efficient way to establish connections between applications and interaction devices rather than to integrate each device individually. A common approach to provide such interconnectivity is the introduction of abstraction layers~\cite{Ohlenburg:2007:DDA:1766311.1766369,Jiang,Scholliers:2010:MDM:1935701.1935712,Chmielewski:2012:AAS:2405172.2405177,VGH12,Konig20108609394} that decouple application logic from I/O handling. This is especially relevant for VR~\cite{Hachet,Taylor:2001:VDN:505008.505019,970514}, where novel devices are introduced at a high rate.

\begin{figure}[!b]
\centering
\includegraphics[width=1.0\columnwidth]{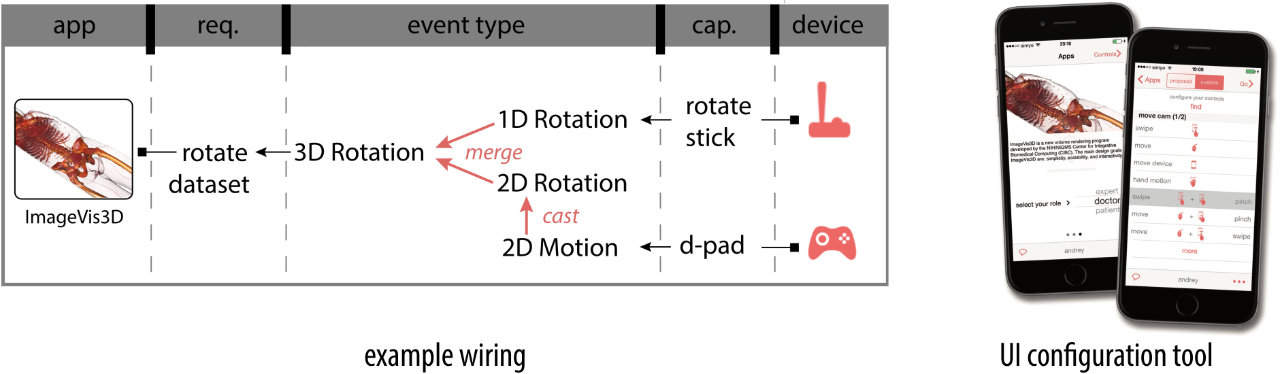}
\caption{The introduced operators allow substituting complex application requirements by combining multiple devices. For instance, rotating the dataset can be achieved by a combination of the d-pad of a gamepad and the stick rotation of a joystick. Users choose between such wiring possibilities via a mobile configuration frontend.}
\label{fig:masterui}
\end{figure}

This section provides a pointer to \textit{MorphableUI}~\cite{muiold,mui}---a multimodal system that transports interaction data between devices and applications. Our approach is based on a taxonomization~\cite{Kin:2012:PMG:2207676.2208694,Ruiz:2011:UMG:1978942.1978971,Dachselt:2006:STM:2386021.2386035,Card:1990:DSI:97243.97263,Mackinlay:1990:SAD:1456610.1456612} of event types, application requirements, and device capabilities. In short, each application relies on our API to state its requirements, and each device exposes its capabilities that describe the type of generated or processed interaction events (cf. \FG{fig:masterui}). 

A token-based hypergraph algorithm is applied to generate iterative solutions that fulfill the application requirements given the currently available devices. As depicted in \FG{fig:hypergraph}, our algorithm allows us to merge, split, and cast device capabilities, e.g., it is possible to combine a 2D swiping input on our phone with two arrow keys on a keyboard to trigger a 3D motion event inside the application. During runtime, our distributed implementation handles the event propagation and transformation via local and remote networks.

\begin{figure}[t]
\centering
\includegraphics[width=1.0\columnwidth]{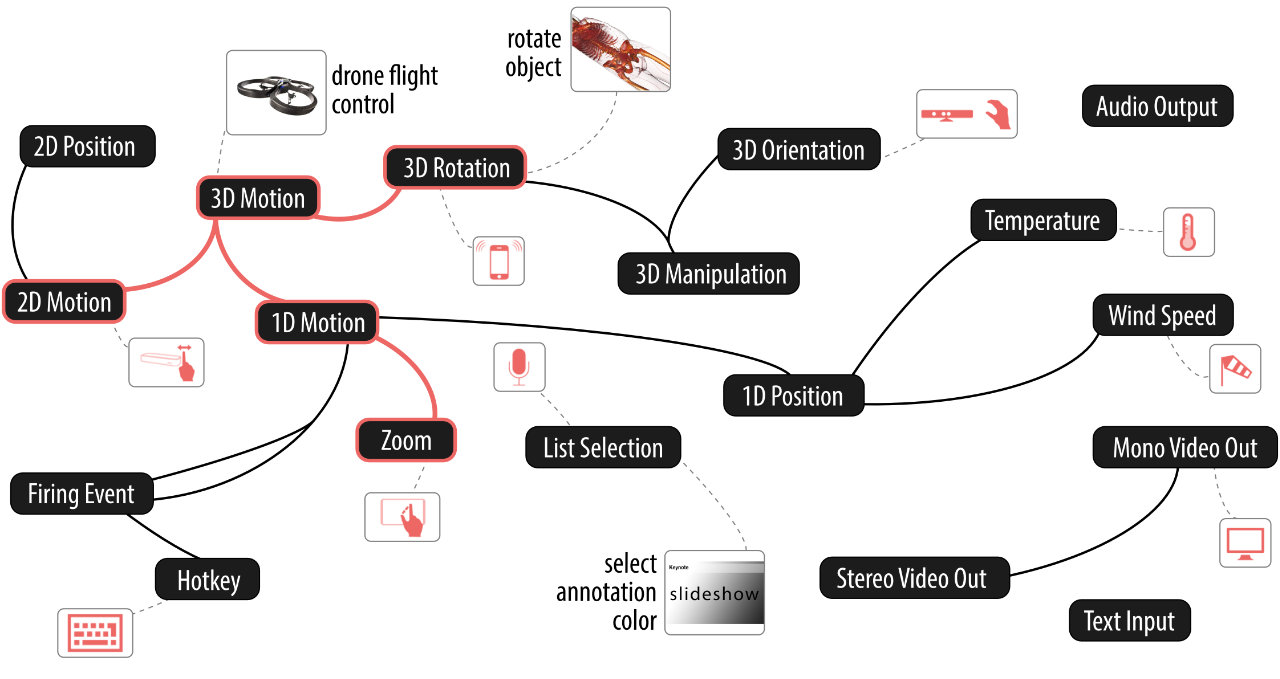}
\caption{An except from the underlying hypergraph model. Event types are captured as vertices, whereas hyperedges represent the operators, i.e., split, merge, and cast. An iterative algorithm starts at device vertices and traverses the hypergraph until it reaches the requirement vertex. The red lines highlight an example of wiring between a 3D rotation requirement and two merged devices.}
\label{fig:hypergraph}
\end{figure}

Our method removes the complexity of I/O handling from the application and allows the integration of novel devices without any modifications to the application logic. Also, \textit{MorphableUI} introduces a visual configuration tool (cf. \FG{fig:masterui}) that allows users to select and adjust their preferred interaction methods. We can reconfigure these user-tailored interfaces~\cite{ICON,Gajos:2004:SAG:964442.964461,Gajos:2007:AGU:1294211.1294253} during runtime, which emphasizes personalization and, in combination with device exchangeability, brings an additional benefit to rapid prototyping scenarios.

Although our work includes validation through a developer survey, we suggest conducting a detailed study that explores the benefits and drawbacks of such user-tailored interfaces. While some of us enjoy tweaking and personalizing the controls of an application, other users will find themselves overburdened by this additional complexity. Hence, as a next step, we strongly recommend an automated interface generation routine that learns from user behavior and provides a starting point for fine-grained configurations.

\section{Applicability of Our Achievements Outside VR}

This chapter was an excursus into the world of VR interaction techniques. We explored several ways to enrich our virtual stay with meaningful object manipulations and natural user interfaces. In particular, we have discussed physical proxies, virtual inventory systems, anticipatable gestures, and I/O abstraction layers. Similarly to the previous locomotion chapter, we dedicate this discussion section to a critical analysis of the outlined contributions.

More precisely, this section reflects upon the applicability of our work beyond virtual environments. In contrast to locomotion, most of our interaction approaches are transferrable to non-VR applications. Even more, certain ideas, such as audience gestures, were inspired by non-VR challenges, which further underpins the generalisability of such solutions. Hence, to make use of the full potential of our publications, we summarize their value outside VR in the following paragraphs:

\textbf{Self-transforming controller.} The straightforward applications are non-VR first-person shooters. From a technical point of view, there is nothing that prevents us from using the controller in a conventional digital game. However, we predict two potential challenges that we need to tackle. Firstly, our prototype purposefully ignores the overall appeal because players do not see the real controller while being in VR. Hence, the---now visible---difference in appearance between the proxy and its virtual counterpart might lead to a reduced increase in player experience. Secondly, the same circumstance could require a more extensive shift in weight that aligns with the virtual transformation.

\textbf{Inventory Systems.} Despite being an established game feature, we are not aware of any research that targets inventories explicitly. Hence, we suggest reconsidering our taxonomy for non-VR games. Even if we need to adjust the individual building blocks due to altered requirements, our manuscript is still a valuable starting point for further investigations. Ultimately, we suggest the creation of a general taxonomy that covers all modalities, including mixed and augmented realities.

\textbf{Comprehensible Gestures.} In a subsequent collaboration, we utilized our gesture sets in a digital planetarium. The presenter interactively navigates through the universe while giving insights into astrophysics and answering live questions. Our gesture sets greatly facilitate such communication. Hence, our crowdsourced gestures have a high potential for many 3D content presentation contexts outside VR that emphasize clear and anticipatable interaction on the part of the moderator.

\textbf{MorphableUI.} Many I/O abstraction layers, such as \textit{VRPN}~\cite{Taylor:2001:VDN:505008.505019}, were created with VR in mind due to the rapidly changing landscape of available devices. In contrast, our hypergraph approach allows for applicability beyond VR-specific setups. The transformation of interaction events facilitates the usage of non-VR devices, such as mobile phones, in VR scenarios, and vice versa. Furthermore, the provided device exchangeability can be used not only for rapid prototyping and user-tailored UIs but also for interactions in changing environments that require dynamic, context-dependent interfaces. On the other hand, \textit{MorphableUI} introduces additional overhead in complexity, both for users and developers. Hence, we recommend to carefully weigh up risks and opportunities before integration, because sophisticated desktop software (e.g., Photoshop) with hundreds of different tasks would most likely not benefit from dynamic UI generation.

To conclude, all of the outlined contributions have significant potential outside VR. Indeed, the conducted research needs to be adjusted (e.g., controller) or extended (e.g., inventories). Nevertheless, our essential observation is that VR-motivated research can remain relevant in human-computer interaction, even if the hype surrounding VR will not prove right.

\chapter{Perception}
\label{sec:perception}

In contrast to locomotion and interaction, this chapter deals with a more subtle, yet not less important aspect of VR, namely perception. While the outlined research of the previous chapters was dedicated to improving the status quo of VR, our work on (visual) perception focuses on the question of how users can benefit from the strong points of VR setups. In other words, the following sections explore the advantages resulting from the altered perception in immersive setups. In particular, we take a closer look at two research areas related to perception: (nonhumanoid) body ownership~\cite{vranims1,vranims2,beastly} and visualization~\cite{dd,deadeye1,deadeye2}.

\section{Nonhumanoid Avatars in VR}

Most VR applications represent our alter ego by using a virtual avatar. Given that VR setups offer a high degree of immersion, the bond to this virtual body is usually much stronger than in common desktop scenarios. Put more simply, VR allows us to perceive the virtual representation as our own body. This perceptional phenomenon is known as the \textit{illusion of virtual body ownership (IVBO)}~\cite{slater2010first}, and previous research agrees that VR is an effective medium to induce such experiences~\cite{slater2009inducing,waltemate2018impact}. 

However, the investigated scenarios were limited mostly to human avatars. Our research~\cite{vranims1,vranims2,beastly}, in contrast, focused on nonhumanoid representations, as we assume that this kind of unusual experience bears significant potential for several application areas, such as entertainment or education. For instance, introducing exotic avatars beyond stereotypic knights and wizards is a viable option to create refreshing and engaging player experiences in VR games. In an educational documentary, the usage of animal avatars could help us to understand the behavior of a particular creature better. Hence, the research on nonhumanoid IVBO outlined in this section is driven by the following motivational question: \textit{Is IVBO applicable to nonhumanoid avatars, and, if so, what potential does that phenomenon have for VR applications?}

\subsection{The Illusion of Virtual Body Ownership}

To facilitate the comprehension of our nonhumanoid research, we start by a brief introduction of IVBO~\cite{lugrin2015anthropomorphism}. This phenomenon originates in the effect of body ownership and the experiments on the so-called \textit{rubber hand illusion}~\cite{botvinick1998rubber}: the participant's arm is hidden and replaced by an artificial rubber limb, and stroking both the real and virtual arms creates the illusion of actually owning that artificial limb. After further studies~\cite{tsakiris2005rubber}, researchers proposed a number of models~\cite{tsakiris2010my,ehrsson2007experimental,petkova2008if,lenggenhager2007video} to explain the interplay between external stimuli and our internal body perception. 
In particular, prior work concludes that, apart from visuotactile and sensorimotor cues~\cite{slater2010first,sanchez2010virtual}, the IVBO effect is mainly impacted by visuoproprioceptive cues (perspective, body continuity, posture and alignment, appearance, and realism)~\cite{slater2009inducing,slater2010first,perez2012my,maselli2013building}. 

The IVBO effect in virtual environments~\cite{slater2008towards,banakou2013illusory} was mainly explored with anthropomorphic characters and realistic representations~\cite{lugrin2015anthropomorphism, lin2016need,jo2017impact}. For instance, regarding avatar customization in games, Waltemate et al.~\cite{waltemate2018impact} showed that customizable representations lead to significantly higher IVBO effects. A strong IVBO can produce various changes in (player) behavior~\cite{jun2018full,muller2017through}, resembling the Proteus Effect by Yee et al.~\cite{yee2007proteus}. For instance, studies revealed a significant reduction in racial bias when players embody a black character~\cite{peck2013putting}. Other examples include childish emotions arising from embodying child bodies~\cite{banakou2013illusory} or feeling more stable when having a robotic avatar~\cite{lugrin2016avatar}. Hence, prior work demonstrates that IVBO can be applied to evoke specific feelings and attributes~\cite{kors2016breathtaking}. Similarly, we assume that a strong bond to an animal caused by IVBO can also increase our involvement with environmental issues~\cite{ahn2016experiencing,berenguer2007effect} and our empathy for animals~\cite{taylor2005empathy}.

Researchers have also expressed interest in studying IVBO beyond human morphology. For instance, Riva et al.~\cite{riva2014interacting} posed the following question: \textit{But what if, instead of simply extending our morphology, a person could become something else---a bat perhaps or an animal so far removed from the human that it does not even have the same kind of skeleton---an invertebrate, like a lobster?} If we consider exotic body compositions, as in the case of a lobster that has few properties in common with our human body, the idea of sensory substitution~\cite{bach2003sensory} might play an important role. Related to VR games, we could also consider such substitution mechanisms as playful interactions: e.g., we could replace the echolocation feature of a bat by tactile feedback in a VR game. Given the extreme diversity of real and fictional creatures, it is difficult or even impossible to research IVBO for virtual animals as a whole. Instead, previous research tackled isolated modifications of body parts. For instance, Kilteni et al.~\cite{kilteni2012extending} were able to stretch the virtual arm up to four times its original length without losing IVBO. Normand et al.~\cite{normand2011multisensory} used IVBO to induce the feeling of owning a more massive belly than in reality. As a first step toward generalization, Blom et al.~\cite{blom2014effects} concluded that strong spatial coincidence of real and virtual body part is not mandatory to produce IVBO.

Individual animals, such as scorpions or rhinos in one of our studies, have additional body parts that players might want to control. In this respect, prior work~\cite{ehrsson2009many,guterstam2011illusion} confirmed that having an additional arm preserves IVBO and induces a double-touch feeling. Steptoe et al.~\cite{steptoe2013human} reported the effects of IVBO upon attaching a virtual tail-like body extension to the user’s virtual character. These findings are relevant for a plethora of real and fictive nonhumanoids, such as dragons. The authors also discovered higher degrees of IVBO when we synchronize the tail movement with the real body.


Naturally, we need a way to measure and compare the IVBO effect in order to investigate whether and how this phenomenon influences our experience in VR. In this regard, the recent work by Roth et al.~\cite{roth2017alpha} introduced the \textit{alpha IVBO questionnaire} based on a mirror scenario. The authors suggest acceptance, control, and change as the three factors that determine IVBO. In our initial experiments, we administered the proposed questionnaire as we were curious to see how it performs for animal avatars. Subsequently, we relied on this questionnaire in all our IVBO experiments to maintain comparability throughout our study results.

\subsection{VR Animals: Applicability of IVBO Beyond Humans}

The first part of our research~\cite{vranims1,vranims2} investigated possible control mechanisms for nonhumanoid avatars and the related levels of IVBO. In general, body ownership requires as much sensory feedback as possible to induce proper levels of IVBO. However, providing such cues is challenging for nonhuman characters because there is no straightforward mapping of controls---animals come in various shapes and postures. For instance, bats share a human posture and skeleton but have scaled arms or legs, i.e., they differ in terms of proportions. Tigers and dogs have an almost human skeleton, including the same number of limbs, but they walk on all fours. Other species, such as a spider, show a completely different skeleton and differ in the limb count. Our publication suggests the following control approaches for animal avatars to cover these different degrees of anthropomorphism:

\begin{figure}[b]
\centering
\includegraphics[width=1.0\columnwidth]{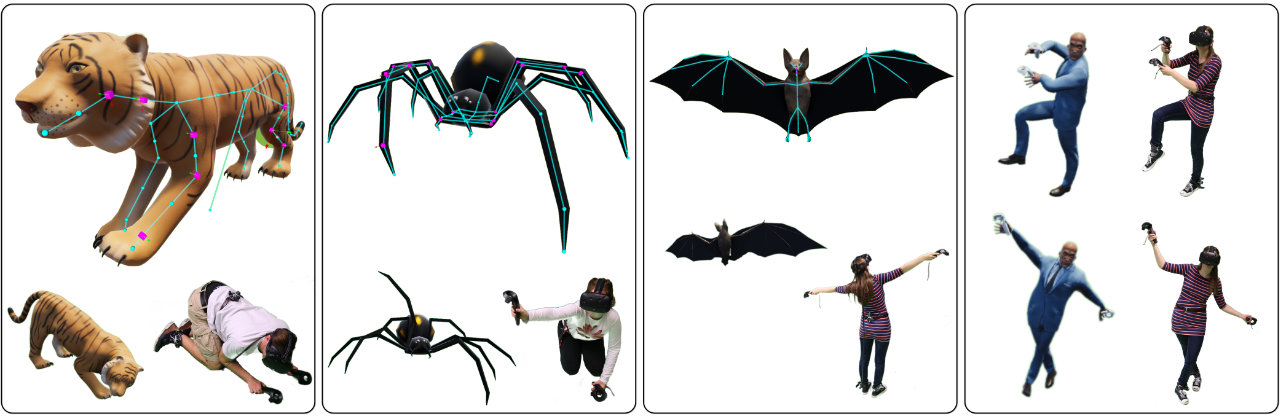}
\caption{Three virtual animals, their controls in full-body tracking mode, and the human avatar that was used as the reference for our IVBO comparisons.}
\label{fig:mapping}
\end{figure}

\textbf{First-person full-body tracking.} The posture of the user is mapped 1:1 to the whole virtual body (cf. \FG{fig:mapping}). In this mode, being a tiger implies that we have to crouch on the floor. From a technical perspective, this approach usually requires an additional tracking of our hip and ankles, which we can easily achieve with mainstream devices, such as Vive trackers~\cite{vive}.

\textbf{First-person half-body tracking.} For particular creatures, a full-body mapping might be too exhausting. Therefore, we designed an alternative that allows us to remain in an upright position while our lower body is mapped to all of the animal’s limbs. In the case of a tiger, each of our legs corresponds to two of the animal’s pawns. Hence, we preserve the sensory feedback and keep the physical effort minimal comparted to full-body tracking.

\textbf{Third-person approaches.} In some applications, we control our avatars from a third-person perspective, i.e., we look over the shoulder of our virtual representation. However, this perspective is challenging when we turn around in VR. There are two ways to handle such rotation. We can either rotate ourselves, i.e., the camera, around the avatar, or vice versa. Camera rotation results in a virtual translation that does not correspond to our physical movement, which, as we remember, is a potential cause of cybersickness. Rotating the avatar can be done in various ways, e.g., by simply sliding the animal sideways around us or by using an agent-like behavior that tries to reposition the animal via natural avatar movement.

\begin{figure}[t]
\centering
\includegraphics[width=1.0\columnwidth]{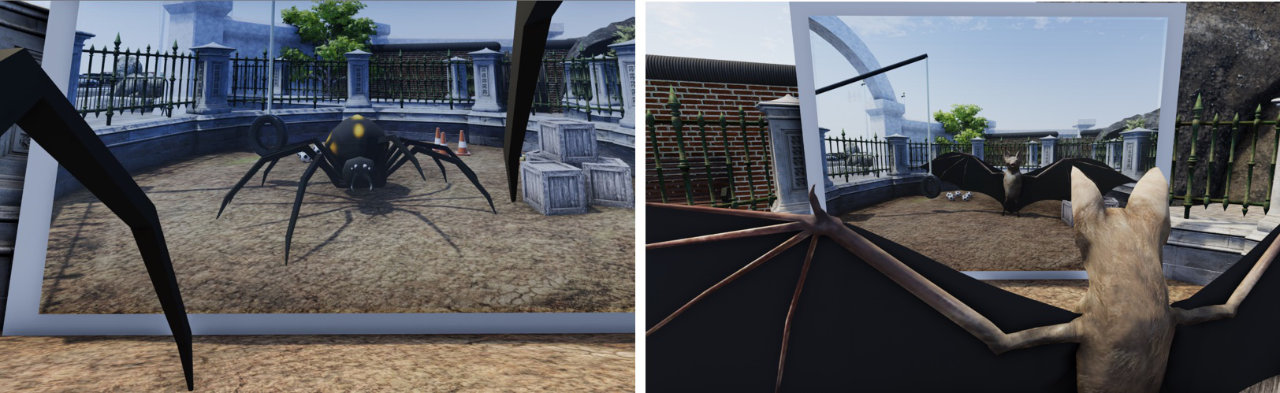}
\caption{The mirror scenario used to assess IVBO for nonhumanoid avatars in first-person (left) and third-person (right) modes.}
\label{fig:mirror}
\end{figure}

We relied on a mirror scenario (cf. \FG{fig:mirror}) to measure the IVBO effects of our control approaches for a tiger, a bat, and a spider. We chose the animals such that they differ from humanoids in IVBO-critical domains, i.e., shape (bat), skeleton (spider), and posture (tiger, spider). With this, we gathered the following insights:
\begin{itemize}[noitemsep]
\item IVBO works for animals and even outperforms humanoid avatars in certain cases (e.g., bat)
\item similarly to human avatars~\cite{galvan2015characterizing}, first-person modes for animals outperform third-person approaches regarding IVBO
\item half-body approaches are a compromise between IVBO and exhaustion: they reduce fatigue for non-upright animals without a noticeable sacrifice of IVBO
\item users enjoy the superhuman abilities (e.g., flying) that come along with a certain animal or an additional body part.
\end{itemize}


\subsection{Beastly Escape: Nonhumanoid Player Experience}

\begin{figure}[!b]
\centering
\includegraphics[width=1.0\columnwidth]{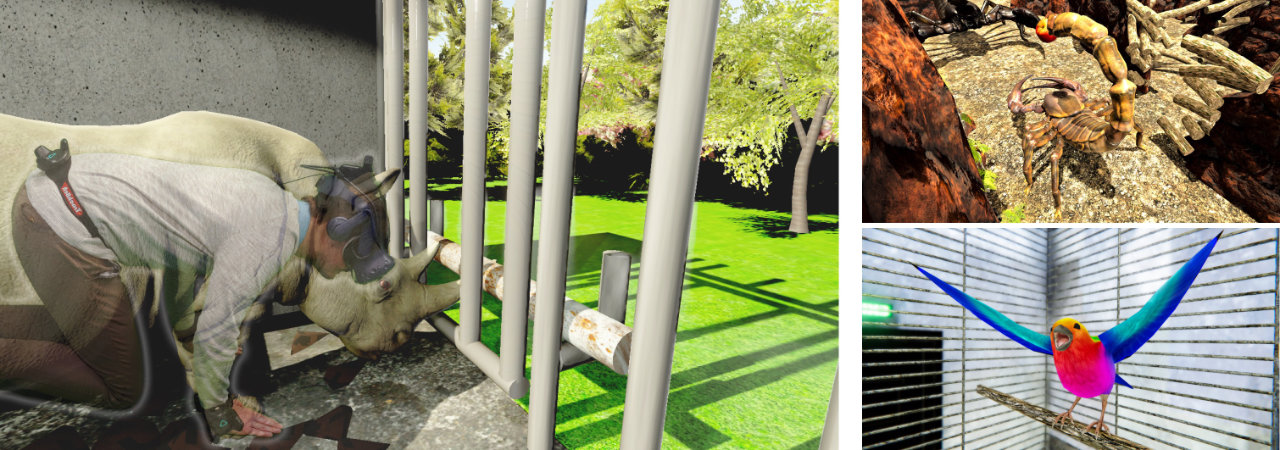}
\caption{Our beastly escape rooms revealed that players enjoy the control over additional body parts that allow novel interactions and enable superhuman abilities.}
\label{fig:escape}
\end{figure}

It is one thing to be aware of nonhumanoid IVBO in virtual environments, but it is another matter to apply this phenomenon in favor of the user---or, in our example, the player. Unfortunately, there are very few studies on creature embodiment in VR, which makes it difficult for game designers to predict whether and how players will perceive animal avatars. As only a few games have touched upon this topic, best practices and design guidelines for such avatars are also lacking.

We contribute to this topic by an in-depth exploration of the design space, the benefits, and the limitations of animal avatars in VR games. Prior work indicates that such research should not be overgeneralized, because animals vary significantly among themselves: they differ in posture, their (loco)motion, and often have entirely different skeletons. Hence, in our work~\cite{beastly}, we decided to maintain a clear focus on a few specific representatives to gather sufficient knowledge regarding how we can embed such avatars in a gaming context.

\begin{figure}[t]
\centering
\includegraphics[width=1.0\columnwidth]{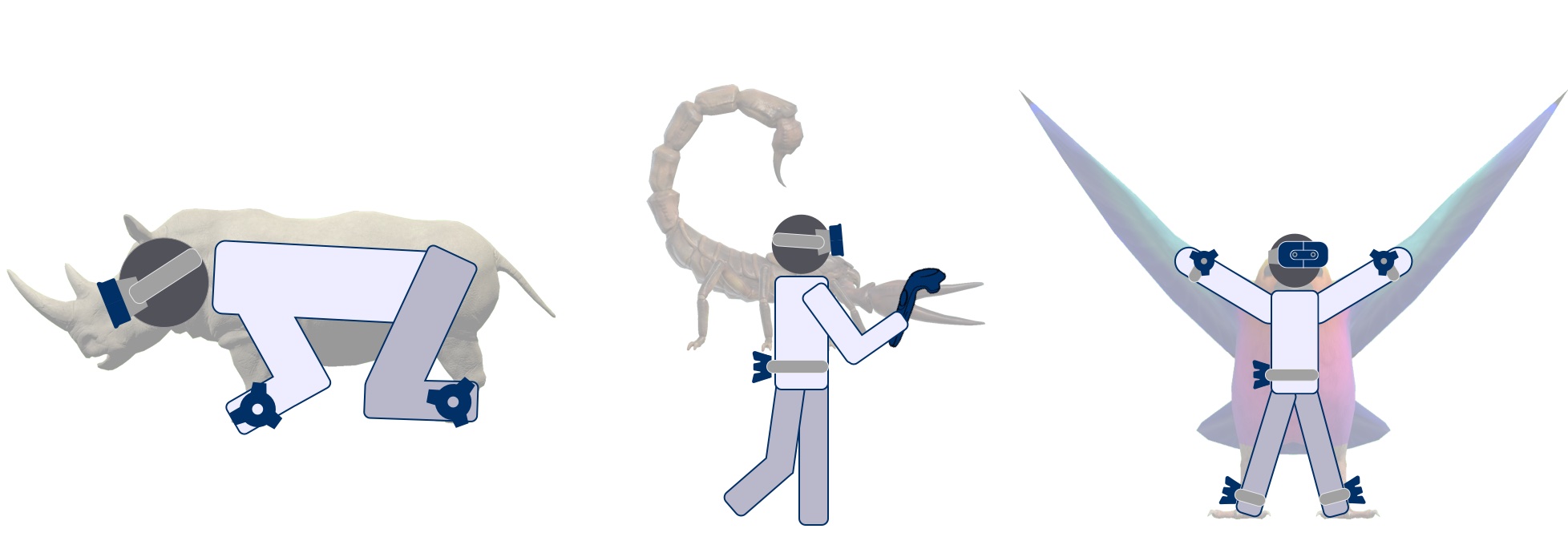}
\caption{The control approaches utilized in our example games. \textbf{Rhino} (full-body mapping): players have to stay on all fours; head movement controls the horn. \textbf{Scorpion} (half-body mapping): players remain in an upright posture and use the controllers to open and close the claws and to initiate a tail strike. \textbf{Bird} (full-body mapping): players use their virtual wings to fly and create gusts of wind.}
\label{fig:beastly}
\end{figure}

In particular, our publication presented the game creation pipeline for escape room games involving three types of animals: a rhino, a bird, and a scorpion (cf. \FG{fig:escape}). The escape room genre was chosen to allow locomotion via natural walking, which removes the need for additional navigation techniques, such as teleportation. As depicted in \FG{fig:beastly}, each game focuses on a different control approach and equips players with a ``superhuman'' skill that is typical for the respective animal. For instance, in the role of a bird, players have to use their virtual wings to fly and to create gusts of wind for object movement. Being a rhino allows players to use the horn for tricky object interactions, such as lifting and removing a lock through the cage bars. A scorpion offers even more unique interactions: in our game, players can use the tail and claws to cut their way through a labyrinth and defeat an end boss.

Our publication provides a detailed discussion of the underlying decision-making process and reports an evaluation of the resulting games concerning IVBO and the overall player experience. We summarize the key messages as following:
\begin{itemize}[noitemsep]
\item IVBO correlates with player experience and is an important factor in nonhumanoid VR avatar design
\item games should emphasize the animal’s specific characteristics and abilities
\item players have no trouble controlling additional and exotic body parts
\item only  directly controllable body parts should be visualized
\item controls (e.g., full-body vs. half-body) should be designed based on the game-related animal abilities and the target audience.
\end{itemize}

Our experiments demonstrated that, by a smart choice of avatars, VR games could allow us to collect impressions and experiences that would not be possible or would be far less engaging in a less immersive setup. To put it another way, nonhumanoid avatars profit substantially from the current VR technology. One reason is the correlation between IVBO, presence, and game enjoyment. Hence, as follow-up research, we suggest disentangling these relations in detail. We assume that further exploration will encourage researchers and practitioners to consider IVBO as a helpful tool that allows the creation of novel and engaging virtual experiences.

Furthermore, we propose to study nonhumanoid IVBO beyond games. In particular, this phenomenon might have an impact on our empathy for animals and nature in general, or we could use it as an in virtuo exposure method~\cite{carlin1997virtual,hoffman2003interfaces,garcia2002virtual,bouchard2006effectiveness,hoffman1998virtual,bouchard2014advances} to combat specific animal-related fears, such as arachnophobia.


\section{Preattentive Highlighting in VR Visualizations}

Animal avatars were just one example of how we can benefit from perception phenomena in VR. In this section, we continue the exploration of such phenomena and take a closer look at the area of visualization---an application domain that is well aware of the advantages that VR has to offer~\cite{kratz2006gpu,shen2008medvis}. 

Our primary contribution is \textit{Deadeye}~\cite{deadeye1}---a preattentive highlighting technique for visualizations, i.e., an approach that allows us to notice an object of interest within a split second. The underlying idea is easy to explain: we highlight an object of interest by rendering it for one eye only. Technically, this is easy to achieve with an HMD, which makes this approach an ideal candidate for VR visualizations. The following sections provide a brief insight into the related research in the following manner: Firstly, we assess the status quo of visualization literacy~\cite{dd}. Secondly, we lay a foundation by establishing \textit{Deadeye} for 2D visualizations~\cite{deadeye1}. Thirdly and finally, we transfer our approach to virtual environments and integrate \textit{Deadeye} into real-world VR visualizations~\cite{deadeye2}.


\subsection{Dear Data: Can Everyone Visualize?}

In the course of this synopsis, we often regarded VR technology as something mainstream, i.e., as something that has an impact on our daily life. One particular reason was the overarching application area, i.e., most of our contributions can be used for entertainment purposes. In contrast, visualization---and visualization creation in particular---is usually considered a ``serious'' application area limited to experts.

Nevertheless, the demand for people being capable of creating meaningful and engaging visualizations outgrows the offer rapidly. Accordingly, the skills to understand and generate visual representations are more crucial than ever---skills that we often subsume under the general term \textit{visualization literacy}~\cite{lee2016vlat}. Hence, before tackling the particular issue of visual highlighting, this section explores how ready the mainstream is for visualization and how we can improve this status quo through education.

Our particular contribution~\cite{dd} is an explorative study on the visualization abilities of novices and a course design that encourages the creation of truly engaging visualizations. The book \textit{Dear Data}~\cite{lupi2016dear} motivated our course. The visualizations in that book were created by composing visualization knowledge and creativity, and we wanted to know if students could be motivated to produce similar results in terms of comprehension~\cite{lankow2012infographics} and engagement~\cite{Bateman:2010:UJE:1753326.1753716,6634103}. Therefore, we emphasized \textit{design thinking}~\cite{Pusca:2018:DTAPS, DMJ:DMJ12001,Goldschmidt:1994:VDTKA,Cross:1996:ADA,Brooks:2010:DDEFCS,Razzouk:2012:WDTI} and hands-on exploration of the visualization space~\cite{roberts2016sketching} to prevent students from any kind of tunnel vision that we often attribute to visualization novices.

\begin{figure}[t]
\centering
\includegraphics[width=1.0\columnwidth]{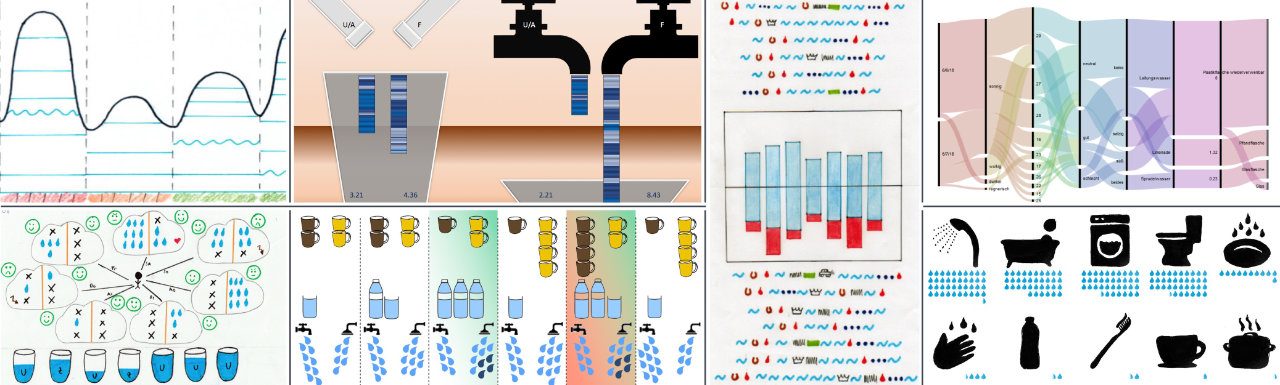}
\caption{Example visualizations on the topic ``water''. Novices experimented with different approaches: day-by-day visualizations vs. aggregations, digital vs. analog, amount of tracked attributes, visual clutter, and differing topic interpretations.}
\label{fig:deardata}
\end{figure}

\begin{figure}[!b]
\centering
\includegraphics[width=1.0\columnwidth]{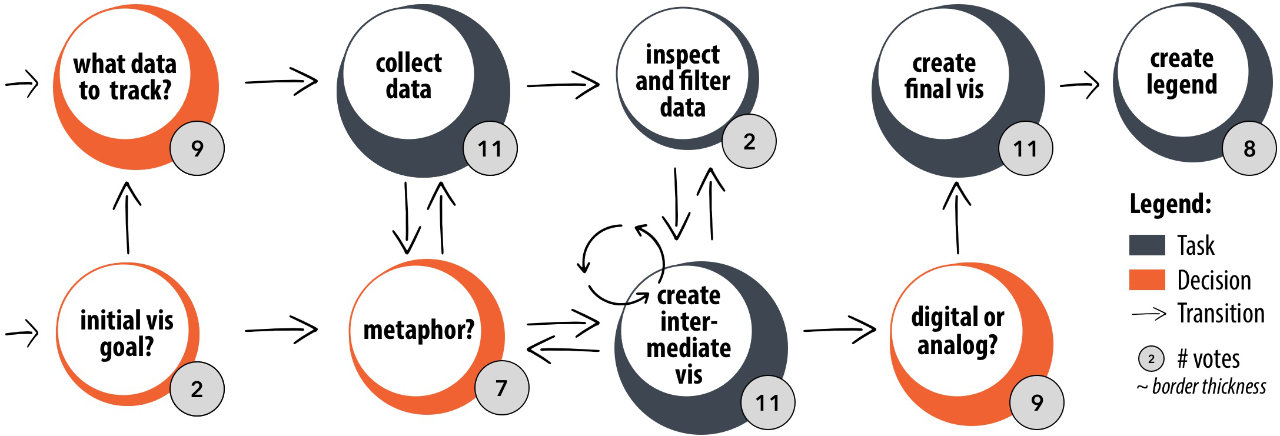}
\caption{ The nonlinear, iterative visualization creation approaches of our students.}
\label{fig:pipe}
\end{figure}

During the course, our participants had to track certain data each week and create a meaningful visualization at the end of the week (cf. \FG{fig:deardata}). Interviews and online surveys accompanied our course in order to collect insights into the thinking process and pain points of visualization novices. Apart from the extracted creation process, as shown in \FG{fig:pipe}, we outline the following key observations:
\begin{itemize}[noitemsep]
\item design thinking motivates novices to experiment with a broader range of visualization methods
\item novices often skip the data analysis step and go straight to visualization
\item novices tend to ignore the aspect of memorability ~\cite{6634103,Bateman:2010:UJE:1753326.1753716} 
\item novices perceive collaborative visualization as more complex and less fun
\end{itemize} 
To summarize, the enrichment of traditional teaching by basic design thinking principles (e.g., divergent thinking, brainstorming) encourages novices to explore and learn a plethora of visualization techniques without falling into a tunnel-vision pattern. We suggest that such novice-oriented approaches can increase the overall visual literacy of the broad masses, which, in turn, is also beneficial for VR visualizations---be it on the part of consumers or producers.

\subsection{Highlighting and Preattentive Visual Features}

As we have seen in the example of \textit{Dear Data}, visualization is an indispensable part of modern communication. Designing comprehensive visualizations requires a deep understanding of how our visual perception works~\cite{NOTON1971929,itti2001computational,yarbus1967eye}. In other words, making efficient use of particular visual characteristics helps us to create visualizations that excel in their usability and performance. 

\begin{figure}[!b]
\centering
\includegraphics[width=1.0\columnwidth]{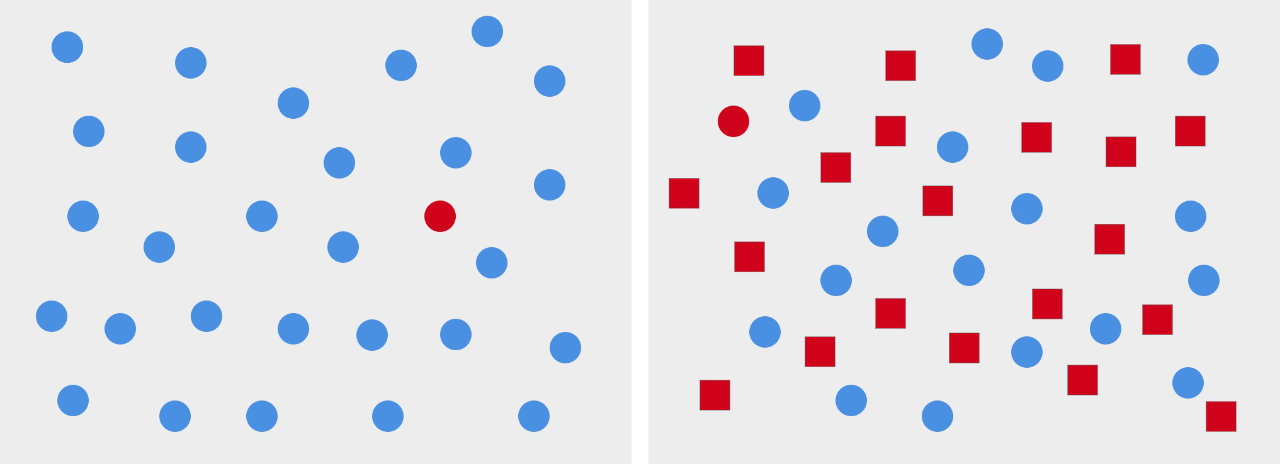}
\caption{Left image: the target object is a red circle among blue distractors and can be recognized preattentively. Right image: the target object is also an outlier -- either a blue square or a red circle (conjunction search). We have to search each object in a serial fashion to find the target, i.e., no preattentive processing is possible.}
\label{fig:preattentive}
\end{figure}

The following sections focus on one aspect of visual perception, namely highlighting, and its interplay with VR technology. Highlighting allows us to draw and guide~\cite{ware2012information,Hall:2016:FEI,Borji:2013:SAV} the attention of users to a particular object of interest. One well-known example is the search function of a web browser: it uses color to highlight the occurrence of a query, which allows us to locate the results instantly. A more advanced example is a medical visualization that utilizes flickering to highlight suspicious cells or tissue and helps doctors with exploring the data. Cues such as color, flickering, shape, size, and motion are examples of so-called \textit{preattentive} visual features~\cite{Healey:2012:AVM:2225054.2225226}. Our visual system can detect such features in a glance, i.e., before our eyes initiate a saccadic movement. Consider the example in \FG{fig:preattentive}: looking at such an image for a split second would suffice to tell whether or not there was a red circle among blue ones. Since a saccade usually needs about 200-250 ms~\cite{Healey:2012:AVM:2225054.2225226} to initiate, researchers utilize that threshold to determine if a cue is preattentive.

One important property of preattentive cues is that they perform equally well with an increasing number of distractors. Those features are processed in parallel by our visual system and are not searched serially. That property is crucial when we revise our example with a full-page text search or the exploration of a huge medical dataset. Hence, advances in the exploration of preattentive features can provide substantial benefits~\cite{Waldner:2014:AFG,Suh:2002:PPA:503376.503422,Cole:2006:DGM:2383894.2383942,Alper:2011:SHG:2068462.2068634,Gutwin:2017:PPI:3025453.3025984} to visualization researchers and practitioners.

\subsection{Deadeye: Dichoptic Presentation for Highlighting}

\begin{figure}[!b]
\centering
\includegraphics[width=1.0\columnwidth]{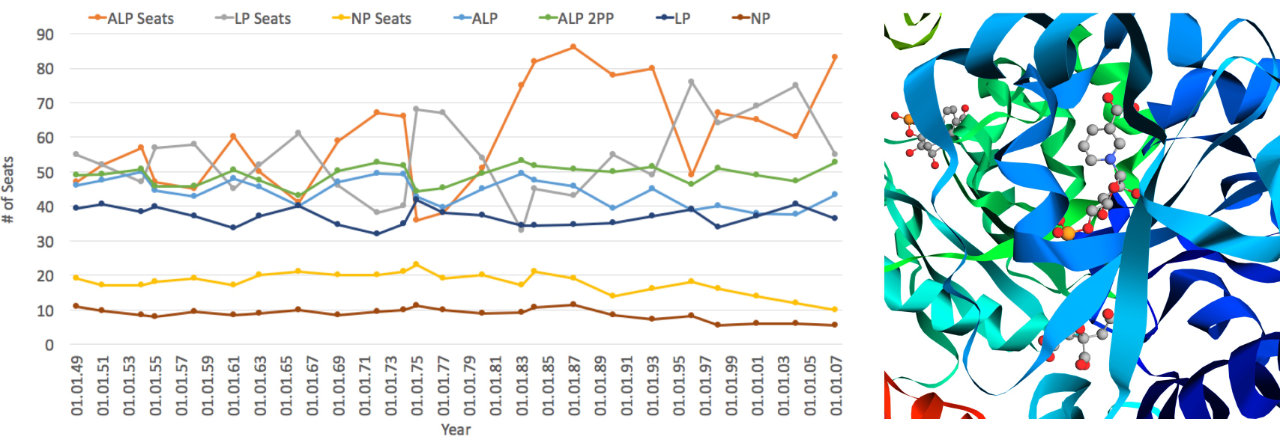}
\caption{Applicability of \textit{Deadeye}: highlighting of lines in a line chart (left) and visual storytelling in a scientific visualization of chemical reactions (right).}
\label{fig:deadeye1}
\end{figure}

The core idea of our work on \textit{Deadeye}~\cite{deadeye1} is to highlight an object by hiding it for one eye. We refer to such a principle when each eye is exposed to a different stimulus as \textit{dichoptic presentation}. In general, that difference in stimuli leads to binocular rivalry~\cite{logothetis1996rivalling, blake1989neural, friedenberg2012visual, alais2005binocular, Paffen2011}, i.e., our vision system enters a context-switching mode that allows us to perceive both monocular images alternately instead of experiencing a superimposition. Whether or not we can perceive binocular rivalry in a preattentive manner has been discussed in several prior works~\cite{wolfe1988binocularity,de1974binocular,teller1967brightnesses,anstis1998nonlinear,formankiewicz2009psychophysics}. The prevailing opinion is that this phenomenon is usually too weak and overridden by more pronounced features~\cite{zou2017binocularity}. Consequently, a dichoptic presentation has only rarely been employed in visualization or for highlighting purposes in general~\cite{Zhang:2012:BSE,Zhang:2014:SBE}. On the other hand, research by Paffen et al.~\cite{paffen2012interocular} and especially the work by Zhaoping~\cite{zhaoping2008attention} has provided further evidence that we should reconsider binocular rivalry as a preattentive cue. 

Our proposed approach has a unique advantage over existing highlighting methods: \textit{Deadeye} does not modify any visual properties of the target and, thus, is particularly suited for visualization applications (cf. \FG{fig:deadeye1}). In contrast, all established cues have to alter the target in one way or another -- be it reshaping, recoloring, or introducing a motion. Such changes in appearance can lead to data misinterpretation. Furthermore, reserving a whole visual dimension, such as color or position, for highlighting is an expensive tradeoff.

We verified our idea using a traditional evaluation approach for preattentive cues. Typically, a series of images are displayed for a short amount of time (100-250 ms), and participants have to decide for each image whether a highlighted object is present or not. A preattentive feature is characterized by a high success rate independently from the number of overall objects, also called distractors, in the image. In addition, we also explored the performance of \textit{Deadeye} in a so-called conjunction search scenario (cf. \FG{fig:preattentive})~\cite{treisman1980feature, treisman1988feature, treisman1986illusory,wolfe1989guided,nakayama1986serial} by combining our technique with color as a second cue. Overall, our results allowed us to draw the following conclusions:
\begin{itemize}[noitemsep]
\item \textit{Deadeye} works preattentively
\item \textit{Deadeye} does not lead to headache or any other physical strain
\item in contrast to the depth cue~\cite{nakayama1986serial}, \textit{Deadeye} can not be processed in parallel when combined with other preattentive features
\item the performance of \textit{Deadeye} decreases with an increasing distance from the focus point, i.e., it is less robust in the peripheral area.
\end{itemize}
The weak spot of \textit{Deadeye} is its dependence on stereo equipment because we have to render different images for each eye. Although the corresponding hardware became a commodity in recent years (e.g., 3D glasses, stereo projectors, 3D TVs), this mandatory requirement leads to an additional effort.  For this reason, a more effective solution would be to transfer \textit{Deadeye} to a natively stereoscopic environment, namely VR, and to apply our approach for 3D visualizations.

\subsection{Deadeye VR: Going Beyond 2D Visualizations}

As a follow-up to our original contribution, we explored \textit{Deadeye} as a highlighting technique for visualizations in VR~\cite{deadeye2}, because such stereoscopic scenarios support dichoptic presentation out of the box. There are manifold reasons for visualization in VR, such as a better understanding of spatial relationships~\cite{schuchardt2007benefits} or the increased presence. Hence, we also need robust and intuitive highlighting techniques. While specific preattentive cues, such as color, are not affected by the transition to VR, temporal approaches, such as flickering, often interfere with aliasing caused by constant micromovements in VR. To put it another way, establishing \textit{Deadeye} as a highlighting method in VR without occupying any additional visual dimension offers significant benefits for the visualization community, as depicted in \FG{fig:deadeyevr}.

\begin{figure}[t]
\centering
\includegraphics[width=1.0\columnwidth]{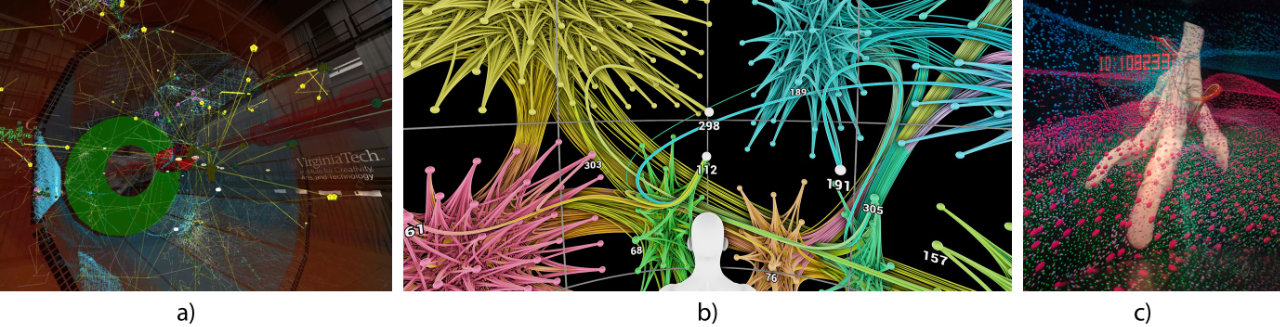}
\caption{Examples of VR visualizations that can benefit from our contribution. (a) Educational visualizations of particle physics~\protect\cite{duer2018belle2vr}: \textit{Deadeye} can be used to capture and guide the attention of the students. (b) Immersive graph visualizations~\protect\cite{kwon16imsv}: utilizing \textit{Deadeye} during user interaction to highlight the selected vertices and edges. (c) Dinosaur track formation~\protect\cite{novotny2019developing}: emphasizing 3D pathlines of interest in unsteady flow visualizations.}
\label{fig:deadeyevr}
\end{figure}

\begin{figure}[!b]
\centering
\includegraphics[width=1.0\columnwidth]{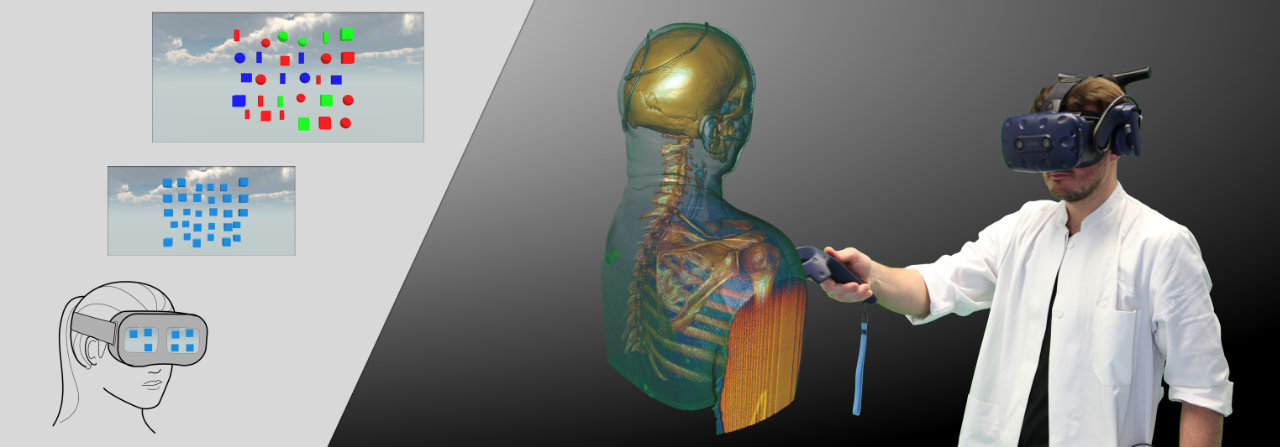}
\caption{Left: our experiments in VR with homogeneous and heterogeneous distractors. Right: VR volume rendering of medical datasets as our evaluated application scenario for \textit{Deadeye}.}
\label{fig:deadeye2}
\end{figure}

We could not assume the applicability of our technique in VR as a given fact. Firstly, dichoptic presentation is a rather subtle cue that might be overridden by more pronounced features~\cite{zou2017binocularity}. Secondly, our method suppresses binocular disparities for the target object. However, our vision utilizes the binocular disparity generated by the horizontal offset of our eyes to gather depth information, which forms the basis for our stereo perception~\cite{julesz1960binocular,julesz1971foundations,Caziot:2015:SOM,marr1976cooperative,marr1979computational}. Therefore, we performed an in-depth evaluation, which revealed the following results:
\begin{itemize}[noitemsep]
\item the preattentiveness of \textit{Deadeye} is preserved in VR
\item \textit{Deadeye} performs robust under heterogeneous conditions, i.e., when distractors vary in protruding properties such as color or shape (cf. \FG{fig:deadeyevr})
\item depth perception for highlighted objects is still possible due to occlusion geometry~\cite{tsirlin2012vinci} and multi-perspective observation~\cite{shimojo1988occlusion}.
\end{itemize}

A second contribution of our publication is an example integration of \textit{Deadeye} into VR volume rendering~\cite{kratz2006gpu,shen2008medvis}, as depicted in \FG{fig:deadeye2}. Along with a GPU-based implementation outline, the manuscript exposes a qualitative survey that demonstrates the benefits and limitations of our approach. According to our results, \textit{Deadeye} is particularly advantageous in non-greyscale scenarios and, in contrast to temporal approaches, does not suffer from typical VR-related issues such as aliasing. Furthermore, our participants noticed that the highlighted object could be faded in or out depending on focus. Simply put, concentrating on something behind the target allows us to suppress the \textit{Deadeye} target completely and see through it. Hence, we suggest investigating this multistable perception phenomenon in detail, as it might have use-cases beyond highlighting, be it for visualizations or VR applications in general.

\section{Concluding Comments on Perception}

The approaches outlined in this chapter slightly differed from our research on locomotion and interaction. In the previous chapters, our contributions were mainly motivated by a shortcoming of current VR setups, be it the limited walking space in a room-scale VR environment or the missing physical feedback of a virtual object. Hence, we proposed methods to improve the status quo of VR by making virtual environments easier to travel and more interactive overall.

In contrast, our methodology concerning perception followed the principle of amplifying the strong points of VR setups. Thanks to the high degree of immersion that VR has to offer, we were able to apply the illusion of virtual body ownership to nonhumanoid avatars, which would be rather hard (or even impossible) to achieve in a desktop environment. Furthermore, we utilized a dichoptic presentation to enhance visualizations with a preattentive highlighting method that does not modify any visual properties of the target object. Hence, visualizations in VR benefit from \textit{Deadeye} out of the box, i.e., without the additional hardware requirements of a typical desktop scenario.


We covered only a tiny fraction of (visual) phenomena that are possible in a virtual environment. Nevertheless, again, the bottom line is that such investigations aim at taking advantage of VR here and now, which further increases the application possibilities of immersive setups. Hence, as researchers, we have to ask ourselves whether and how we can operationalize the search for such phenomena. From our perspective, it is more promising to start in VR and to explore all nuances of virtual experiences, instead of porting existing approaches from desktop to VR, as is often done---with little success---with digital games.

This ``VR first'' paradigm, combined with divergent thinking (cf. our work on \textit{Dear Data}), is something that can not be emphasized often enough concerning our perception in VR. With that, we do not only mean the exploration of real-world inspired perception phenomena: certain \textit{unrealistic} perceptions, such as being an animal, are only possible in VR, which further underpins the diversity and advantageousness of such setups.

\chapter{Conclusion}
\label{sec:conclusion}

Throughout this synopsis, we explored three different areas of VR research: locomotion, interaction, and perception. We began by classifying the different movement approaches in VR. We outlined stationary approaches, such as gamepad or gaze-based controls, and summarized the benefits of walking-inspired techniques, such as walking in place or redirected walking. We then focused on virtual body resizing as one viable option to bring natural walking into our living rooms. Based on our \textit{GulliVR} and \textit{Outstanding} publications, we exposed how to use player rescaling to increase presence and spatial orientation while removing the risk of cybersickness.

Our stay in a virtual environment seldom remains limited to just walking---hence, in the next step, we presented various ways to enhance and facilitate interaction in VR. We proposed a self-transforming controller to increase presence and player enjoyment in VR games and embedded this topic in a more general discussion about object representations and inventories. Apart from these object-related contributions, we investigated gestures as an essential entity in interaction research. We emphasized the audience and created 3D gestures that are easy to understand and to predict by an observer, be it during a presentation, a software demo, or for entertainment purposes. Finally, we introduced \textit{MorphableUI}, a distributed hypergraph-based system that facilitates the integration and usage of diverse I/O modalities in and outside VR, such as the aforementioned physical proxies and gestures.

In the case of interaction and locomotion, we mainly dealt with the shortcomings of VR. For instance, we addressed issues such as the limited space for walking or the (mostly) missing haptics. Our contributions in the field of perception emphasize the strong points of immersive setups. We have transferred the illusion of virtual body ownership to nonhumanoid avatars and exploited this phenomenon to create engaging and novel gaming experiences with animals in the leading role. As a further example, we introduced \textit{Deadeye} and demonstrated how to utilize the dichoptic presentation capability of immersive setups for preattentive highlighting in visualizations. In particular, our technique allows guiding the attention to an object of interest with the unique benefit of not modifying any visual property of the target.

With these accomplishments in mind, we dedicate the remaining sections of this chapter to a critical assessment of our contributions and their potential regarding follow-up research. We round off the synopsis by presenting our concluding thoughts on VR research in general and outlining our lessons learned along the way.

\section{Critical Assessment of the Contributions and Possible Future Work}
We outlined 16 manuscripts in this synopsis. It is fair to assume that they differ regarding their overall impact on the status quo of VR and computer science research in general. Assessing the significance of a contribution shortly after the publication date is somewhat speculative, and the rating of own works is always prone to personal bias. Nevertheless, such an initial judgment helps to complete the big picture of this dissertation. 

To maintain focus, we do not discuss each contribution in detail. Instead, for each research area, we outline the---from our point of view---most significant achievement(s), reason our choice, and provide possible follow-up research questions.

\textbf{Locomotion.} Our major achievement in this area is the introduction of the virtual body resizing concept. The award-winning \textit{GulliVR} publication (\textit{CHI Play 2018 honorable mention}) demonstrates how an appropriate adaptation of the virtual eye distance can be used to move at an increased pace without the risk of cybersickness. Although prior research reported that an increased stereo base makes objects appear nearer and smaller~\cite{renner2015influence,van2011being}, we are not aware of any previous attempts to utilize this phenomenon as a viable countermeasure for locomotion-induced cybersickness. Our core idea is applicable beyond games and entertainment, which is reflected by the diversity of citing sources~\cite{g1,g2} and follow-up publications~\cite{abtahi2019m,outstanding1,outstanding2}. We see two possible research directions based on \textit{GulliVR}: Firstly, the dynamic resizing approach has a straightforward use-case in multiscale virtual environments~\cite{argelaguet2016giant,Zhang:2002:SIM:571878.571884,kopper2006design}, be it for urban planning or medical explorations. Secondly, we suggest further experiments on the significantly altered virtual eye separation. We believe that the induced change in the perception of object size and distance is a subtle, yet powerful phenomenon that creates opportunities for several novel VR experiences, such as the miniature world look in the case of \textit{Outstanding}~\cite{outstanding1,outstanding2}.

\textbf{Interaction.} In the respective chapter, we concluded that most of our results are also applicable outside VR. However, as our assessment focuses on VR research, we emphasize our work on self-transforming controllers as an essential contribution in this area. We regard this publication as a significant milestone because of our findings related to weight shifting and weight perception in VR. In contrast to the pre-existing opinion that a physical proxy has to mimic the virtual object as close as possible, our research demonstrated that such proxies could (and should) be significantly lighter and handier. Hence, we have shown that most of the commercially available VR gun controllers are on the wrong path, as they aim to replicate the exact look and feel of the real gun, which is unnecessary and even counterproductive. More importantly, we have determined that a slight shift of weight is sufficient to convey the impression of holding a completely different object (or gun in our case), which paved the way for further experiments on self-transforming controllers~\cite{zenner2019drag}. Since haptics plays a crucial role in our virtual experience, we argue that research on similar morphable controllers significantly advances the status quo of VR. As future work, we suggest a broader exploration of such transforming devices by including additional parameters aside from weight distribution. For instance, we can modify air resistance, as in the case of \textit{Drag:on} by Zenner et al.~\cite{zenner2019drag}, or add a tactile surface representation similar to the \textit{Haptic Revolver} by Whitmire et al.~\cite{whitmire2018haptic}.

\textbf{Perception.} We outline two significant contributions---VR Animals in the entertainment sector and \textit{Deadeye} in the area of visualization. Our work on animal avatars pioneered the concept of nonhumanoid body ownership in VR. In multiple experiments, we confirmed that IVBO applies to avatars that are very different from human beings. These insights allowed us to create novel, engaging virtual experiences with animals in the primary role. Our research promotes the injection of superhuman abilities, such as flying or using a virtual horn, into VR games as a central component of player experience. Although our publications emphasize digital games, we assume that animal embodiment is also a viable enhancement for other domains. In particular, we hypothesize that being an animal in VR increases our empathy for animals and might be applicable as an in virtuo exposure method to combat animal-related fears.

In contrast to this certainly unusual topic, our work on \textit{Deadeye}, an \textit{IEEE VIS 2018 best paper award} publication, is characterized by its simplicity and straightforward advantages for everyday visualizations. The discovery of a novel preattentive visualization method is a rare occasion. Together with the fact that \textit{Deadeye} does not modify any visual property of the target, it is reasonable to assume a high significance of this contribution. The applicability of our method in VR is what brings out the full potential of \textit{Deadeye} because dichoptic presentation comes out of the box in such setups. As a result, we get a (less than) zero-overhead highlighting method that is straightforward to implement while preserving all attributes of the target object. As a next step, we suggest to go beyond visualizations and to evaluate \textit{Deadeye} for graphical user interfaces in VR, as they heavily rely on highlighting and attention guidance and would supposedly benefit from the non-invasiveness of our method.

\section{VR Research and the Lessons We Learned}

Our synopsis began with the question of whether (our) VR research is ``worth it''. In the previous section, we outlined our most significant contributions and their potential impact. However, final publications are not everything---along the way, we encountered several hidden obstacles, failed at particular challenges, and gained scientific maturity. Hence, as the last words of the synopsis, we summarize the essential lessons that we have learned during our VR research.

\textbf{VR first instead of fixes and ports.} We noticed that VR is often treated as an extension of conventional desktop-based computing. We hear people talking about ``porting an application to VR'' or reading changelogs of games stating that a particular element, such as the user interface or locomotion, `` was adapted to VR''. From our experience, this approach of taking an application, rendering it in stereo, i.e., in VR, and then asking oneself how to fix the countless usability issues is worrying, saying the least. Not only that such fixes are difficult or sometimes even impossible to achieve---this patching process leaves a certain after-taste and increases the overall skepticism regarding the maturity of VR. Moreover, how often have we witnessed applications that remained unusable in VR, no matter how many patches were applied? We argue that, as researchers and practitioners, we must distance ourselves from this ``port-and-fix'' mindset. If we really want to create meaningful experiences in VR, we have to embrace this entity as a whole, with all its benefits and drawbacks, which often means starting from scratch. Do not get us wrong here, most principles of human-computer interaction still apply to virtual environments, and that brings us to the next point: 


\textbf{Reinventing the wheel in VR.} For years, VR research is booming. Related conferences, such as the \textit{IEEE VR}, went from meet-ups with a few hundred attendees to multi-track symposiums with thousands of visitors. Conferences on human-computer interaction, such as the \textit{ACM CHI}, dedicate multiple sessions to VR research nowadays. This seemingly rapid progress has both advantages and disadvantages. On the one hand, the increasing number of scientists is, without a doubt, an essential factor for the sustainable success of VR. More workforce results in more ideas and broader dissemination of the respective findings thanks to our lively, vibrant community. On the other hand, we noticed that VR research is occasionally accompanied by some superficiality---instead of diving into the literature on stereoscopic environments from the nineties, we tend to overlook previous efforts and claim novelty just because nobody tried our method before with an up-to-date VR HMD.

Another example of reinventing the wheel is to (re-)publish methods and applications by adding \textit{in VR} to the title. Not all tasks are automatically better in a virtual environment---we would even go so far as to say that most daily use-cases are better off without VR. We should always ask ourselves, whether the VR-induced benefits to our application justify the fact that people have to put on a slightly uncomfortable HMD, potentially experience cybersickness, and lose the advantages of a desktop environment, such as efficient text input. For instance, we needed VR in our animal avatar research, as we built upon the virtual body ownership illusion. Otherwise, a non-VR setup would have enabled such nonhumanoid experiences for a wider audience. This example also introduces the last lesson we want to share:

\textbf{Simulations vs. unrealistic experiences.} Certainly, we can utilize the provided immersion of VR setups for ultra-realistic simulations. We can design our virtual stay to match our real life as close as possible. However, in our opinion, one of the greatest strengths of VR is to deliver \textit{unrealistic} experiences. In VR, we are faced with unique opportunities and activities that are not available to us in the real world. In our experiments, the participants were particularly attracted by various \textit{unrealistic} actions like flying or interacting with a miniaturized world as a giant. Ultimately, we believe that such novel experiences are vital to the success of VR because curiosity is an integral part of human nature.





%
{%
\setstretch{1.1}
\renewcommand{\bibfont}{\normalfont\small}
\setlength{\biblabelsep}{0pt}
\setlength{\bibitemsep}{0.5\baselineskip plus 0.5\baselineskip}
\printbibliography[nottype=online]
\newrefcontext[labelprefix={@}]
\printbibliography[heading=subbibliography,title={Webpages},type=online]
}
\cleardoublepage




\appendix\cleardoublepage

\cleardoublepage

\cleardoublepage
%
\pdfbookmark[0]{Declaration}{Declaration}
\addchap{Declaration}
\label{sec:declaration}
\thispagestyle{empty}

I hereby declare that I have completed my work solely and only with the help of the references I mentioned.

\bigskip

\noindent\textit{\thesisUniversityCity, \thesisDate}

\smallskip

\begin{flushright}
	\begin{minipage}{5cm}
		\rule{\textwidth}{1pt}
		\centering\thesisName
	\end{minipage}
\end{flushright}


\clearpage

\newpage
\mbox{}

\end{document}